\RequirePackage{fix-cm}
\documentclass[smallcondensed]{svjour3}     
\smartqed  
\usepackage{graphicx}
\usepackage{layout}
\usepackage{natbib}
\usepackage{color}
\usepackage[bookmarks=true,bookmarksnumbered=true,bookmarkstype=toc]{hyperref}
\hypersetup{colorlinks=true, citecolor=blue, linkcolor=blue, urlcolor=blue}
    \usepackage[misc,geometry]{ifsym} 
\usepackage{amssymb}
\usepackage{mathtools}
\usepackage{setspace}
\usepackage{multirow}
\usepackage{longtable}
\usepackage{dcolumn}
\usepackage{arydshln}
\usepackage{booktabs}
\journalname{Asia-Pacific Journal of Regional Science}
\def\makeheadbox{\relax}
\begin{document}

\title{Bilateral multifactor CES general equilibrium with state-replicating Armington elasticities\thanks{This material is based upon work supported by JSPS Grant No. 16K00687.
}
}

\titlerunning{Bilateral multifactor CES general equilibrium}        

\author{
Jiyoung Kim \and Satoshi Nakano \and Kazuhiko Nishimura 
}


\institute{
J. Kim \at Institute of Developing Economies, Chiba 261-8545, Japan\\ \email{jiyoung\_kim@ide.go.jp}          
\and
S. Nakano \at The Japan Institute for Labour Policy and Training, Tokyo 177-8502, Japan \\ \email{nakano@jil.go.jp}
\and
K. Nishimura \Letter \at Faculty of Economics, Nihon Fukushi University, Tokai 477-0031, Japan \\ \email{nishimura@n-fukushi.ac.jp}
}

\date{\today}

\maketitle

\begin{abstract}
We measure elasticity of substitution between foreign and domestic commodities by two-point calibration such that the Armington aggregator can replicate the two temporally distant observations of market shares and prices.
Along with the sectoral multifactor CES elasticities which we estimate by regression using a set of disaggregated linked input--output observations, we integrate domestic production of two countries, namely, Japan and the Republic of Korea, with bilateral trade models and construct a bilateral general equilibrium model.
Finally, we make an assessment of a tariff elimination scheme between the two countries.
\end{abstract}
\begin{flushleft}
\end{flushleft}

\section{Introduction}
\label{intro}
Recently, \citet{knn} established a general equilibrium framework comprising multifactor CES production functions with estimated elasticities for each of the industrial sectors; each elasticity is measured by the slope of the regression line between the growths of factor shares and factor prices, which are observed in a set of linked input--output tables.
The present study is intended to extend this framework in such a way as to incorporate substitution between domestic and imported commodities (i.e., Armington elasticity) and to endogenize the international trades for all commodities traded between two countries, namely, Japan and the Republic of Korea.

Armington elasticity is an essential component in trade policy analyses.
Previous work concerning economic assessment of trade liberalization schemes \citep[e.g.,][]{hrt, nakajima, uratakiyota, bokep}
have used computable general equilibrium (CGE) models based on the Global Trade Analysis Project (GTAP) database.
While these models make use of empirically estimated elasticities, the estimates of the elasticities between the aggregated factors, which are essentially based upon time series analyses, tend to show smaller elasticities than anticipated \citep{mdb}.
Notwithstanding, Armington elasticities can be large in light of the indifferences between goods of the same classification but from different countries.
Moreover, previous CGE models are calibrated at a single state (i.e., one-point calibration) to incorporate regionalities, except for the elasticity parameters, which are usually given a priori.

From another perspective, \citet{imf} was concerned with the separability of foreign commodities i.e., the distinction between inter- and intra-group Armington elasticities.
The inter-group elasticity is the elasticity of substitution between a basket of domestic commodities and that of imports as a whole, whereas the intra-group elasticity is the elasticity of substitution between a basket of imports from one foreign country and that from another.
The estimates of inter-group elasticities were larger for intermediate input sectors, whereas the intra-group elasticities were significantly lower.
In the same vein, \citet{feenstra} studied the elasticity of substitution between domestic and foreign goods (i.e., macro elasticity), and between varieties of foreign goods (i.e., micro elasticity) and essentially found the opposite: the micro elasticity was significantly larger than the macro elasticity.

\begin{figure}[t!]
\includegraphics[scale=1]{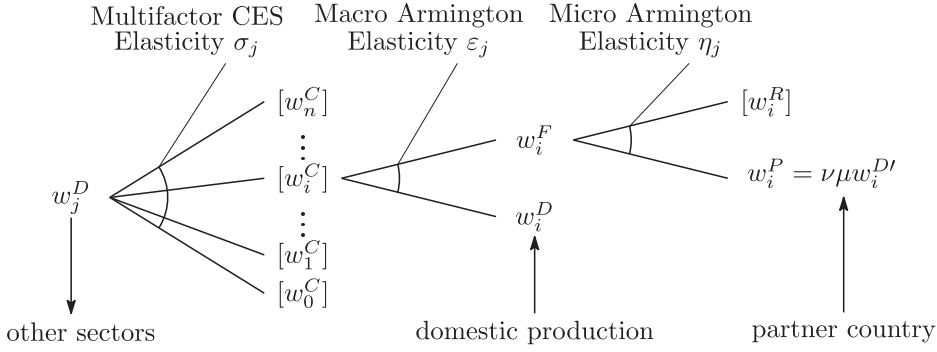}
\caption{Nested structure of macro and micro Armington elasticities.
A foreign commodity price is given by aggregating the partner country's and the rest of the world's (ROW's) commodity prices.  
The compound commodity price is given by aggregating the domestic and foreign commodities' prices.  Finally, the domestic price is given by a multifactor CES aggregator (i.e., unit cost function). 
}
\label{nesting}      
\end{figure}

Our approach differs from those of the previous studies in two aspects.
First, all elasticities are measured based upon published statistics, i.e., linked input--output tables for Japan and for Korea and the UN Comtrade database, and are not adopted from elsewhere.
Second, we construct a model that completely replicates two temporally distant state observations rather than conducting a time series analysis to measure elasticities between aggregated factor inputs, as we are interested in a shorter term and a sector-wide policy implication such as regarding tariff liberalization.
The state-replicating Armington elasticities are measured by two-point calibration.
That is, we measure elasticity that agrees with the two observed domestic--foreign shares in both physical and monetary terms.
Moreover, the elasticities are measured in a two-stage nested structure as illustrated in Fig. \ref{nesting}.

Specifically, we evaluate the compound price of each factor input $w_{i}^C$, in terms of domestic and foreign factor input prices ($w_{i}^D$, $w_{i}^F$) that are observable, via CES aggregation whose macro elasticity replicates the observed domestic-foreign market shares.
We then calibrate the micro elasticity by using $w_{i}^F$ and the partner country's domestic price $w_{i}^{D\prime}$ in order that the observed partner-ROW  market shares are replicated.
In this way and based upon 2000--2005 linked input--output tables for Japan and Korea, we construct a multi-sectoral (395 for Japan and 350 for Korea) general equilibrium model with endogenized bilateral trades, in contrast to the previous studies with limited variety of industrial sectors.

The remainder of the paper is organized as follows.
In the next section, we introduce the basics of the two-point calibration of the CES elasticity parameters, i.e., macro and micro Armington elasticities, and the multifactor CES elasticity estimation by regression.
In Section 3, we apply these protocols using linked input--output tables for Japan and for Korea and the UN Comtrade database.
In Section 4, we integrate domestic and trade models to construct a bilateral general equilibrium model for welfare analysis of trade liberalization.
Section 5 provides concluding remarks.

\section{The Model}
\subsection{Macro Aggregator}
Assume foreign and domestic commodities are, to a certain extent, substitutes with a constant elasticity of substitution (CES).
Then, a composite product price in a country (whose index $i$ is omitted) can be evaluated by a CES aggregator of foreign and domestic commodity prices as follows:
\begin{align}
w^C=\left( \alpha \left(w^D\right)^{1-\varepsilon} + (1-\alpha) \left(w^F\right)^{1-\varepsilon} \right)^{\frac{1}{1-\varepsilon}}
\equiv {U}\left( w^D, w^F \right)
\label{one}
\end{align}
where $w^C$ is the composite price of a commodity in the concerned country, $w^F$ is the price of the imported foreign product (including tariff), and $w^D$ is the price of the domestic commodity.
Here, the share parameter $\alpha \in (0,1)$ and the macro Armington elasticity $\varepsilon$ are subject to estimation.

According to Shephard's lemma, we can obtain the cost share by taking derivatives as follows:
\begin{align}
&s^D=\frac{\partial w^C}{\partial w^D} \frac{w^D}{w^C}=\alpha \left( \frac{w^D}{w^C} \right)^{1-\varepsilon}
&s^F=\frac{\partial w^C}{\partial w^F} \frac{w^F}{w^C}=(1-\alpha) \left( \frac{w^F}{w^C} \right)^{1-\varepsilon}
\label{two}
\end{align}
where $s^D$ and $s^F$ denote the market shares of the domestic and imported commodities, respectively.
One may verify that $s^D + s^F=1$ by taking (\ref{one}) into account.
Below we show that $\varepsilon$ can be measured by two-point calibration using two temporally distant market share observations, namely, the reference market shares $( s_<^D, s_<^F )$ and the current market shares $( s_>^D, s_>^F )$, with the price changes in the domestic $(w_<^D, w_>^D)$ and imported commodities $(w_<^F, w_>^F)$.
Now, according to (\ref{two}), the identities 
\begin{align}
&s_<^D=\alpha \left( \frac{w_<^D}{w_<^C} \right)^{1-\varepsilon}
&s_<^F=(1-\alpha) \left( \frac{w_<^F}{w_<^C} \right)^{1-\varepsilon}
\label{three}
\end{align}
must hold at the reference state and the identities 
\begin{align}
&s_>^D=\alpha \left( \frac{w_>^D}{w_>^C} \right)^{1-\varepsilon}
&s_>^F=(1-\alpha) \left( \frac{w_>^F}{w_>^C} \right)^{1-\varepsilon}
\label{four}
\end{align}
must hold at the current state. By virtue of (\ref{three}) and (\ref{four}), $\varepsilon$ can be solved (two-state calibrated) as follows:
\begin{align}
\varepsilon 
=1-\frac{\ln {s_>^D}/{s_<^D} - \ln {s_>^F}/{s_<^F}}{\ln {w_>^D}/{w_<^D} - \ln {w_>^F}/{w_<^F} }
=1-\frac{\Delta \ln s^D - \Delta \ln s^F}{\Delta \ln w^D - \Delta \ln w^F}
\label{five}
\end{align}
where $\Delta$ is the difference operator, i.e., current value minus reference value.
Also, we may solve for the share parameter $\alpha$ as follows:
\begin{align}
\frac{\alpha}{1-\alpha} 
= \frac{s_<^D}{s_<^F}  \left(\frac{w_<^F}{w_<^D} \right)^{ (1-\varepsilon) }
= \frac{s_>^D}{s_>^F} \left(\frac{w_>^F}{w_>^D} \right)^{ (1-\varepsilon) }
\label{sss}
\end{align}
In this way, we obtain the macro Armington aggregator (\ref{one}) that replicates both the reference and current states specified by (\ref{three}) and (\ref{four}), respectively.
We also note that the compound price $w^C$ will be evaluated assuming (\ref{one}) and thus it is shown in brackets in Fig. \ref{nesting}.

\subsection{Micro Aggregator}
Let us indicate the partner country by $P$ and the ROW by $R$.
Assume that the aggregated foreign import product price $w^F$ (whose commodity index $i$ is omitted) can be expressed as a CES aggregator function of price of commodity imported from the partner country $w^P$ and that from the ROW $w^R$, as follows:
\begin{align}
w^F= \left( \beta \left( w^P \right)^{1-\eta} +(1-\beta) \left( w^R \right)^{1-\eta}  \right)^{\frac{1}{1-\eta}}
\equiv {V}\left( w^P, w^R \right)
\label{six}
\end{align}
where $\beta \in (0, 1)$ is the share parameter and $\eta$ is the micro Armington elasticity, both of which are subject to estimation.
Note that $w^R$ must be evaluated assuming (\ref{six}) with the calibrated parameters, while $w^F$ and $w^P$ are statistically observable.\footnote{As we will be discussing later, the price of the commodity from the partner country $w^P$ will be measured by using the partner country's domestic price $w^{D\prime}$, the relative import barrier factor with respect to the partner country ${\mu}$, and the currency exchange factor $\nu$, i.e., $w^P=\nu {\mu} w^{D\prime}$.}
Hence, the parameters are calibrated according to the two-state observation of the partner country's market share within the commodity's fraction of imports, i.e., $(s_<^P, s_>^P)$.
Notice that $s_<^P + s_<^R = 1$ and $s_>^P + s_>^R = 1$ by definition.

The following identities must hold at the reference state, according to Shephard's lemma applied to (\ref{six}):
\begin{align}
&s_<^P=\beta \left( \frac{w_<^P}{w_<^F} \right)^{1-\eta}
&s_<^R=(1-\beta) \left( \frac{w_<^R}{w_<^F} \right)^{1-\eta}
\label{eight}
\end{align}
Likewise, the following identities must hold at the current state:
\begin{align}
&s_>^P=\beta \left( \frac{w_>^P}{w_>^F} \right)^{1-\eta}
&s_>^R=(1-\beta) \left( \frac{w_>^R}{w_>^F} \right)^{1-\eta}
\label{nine}
\end{align}
By virtue of (\ref{eight} left) and (\ref{nine} left), $\eta$ can be solved (two-state calibrated) as follows:
\begin{align}
\eta
=1-\frac{\ln {s_>^P}/{s_<^P} }{\ln {w_>^P}/{w_<^P} - \ln {w_>^F}/{w_<^F} }
=1-\frac{\Delta \ln s^P }{\Delta \ln w^P - \Delta \ln w^F}
\label{ten}
\end{align}
Also, we may solve for $\beta$ as follows:
\begin{align}
\beta
= {s_<^P}\left(\frac{w_<^F}{w_<^P} \right)^{1-\eta}
= {s_>^P}\left(\frac{w_>^F}{w_>^P} \right)^{1-\eta}
\label{eleven}
\end{align}
Hence, we have the micro Armington aggregator (\ref{six}) that replicates both the reference and current states. 
Also note that $w^R$ will be evaluated by (\ref{six}): 
\begin{align*}
w^R = \left( \frac{\left(w^F\right)^{1-\eta} - \beta \left(w^P\right)^{1-\eta}}{1-\beta} \right)^{\frac{1}{1-\eta}}.
\end{align*}

The in-bound price of the product imported from the partner country $w^P$ is evaluated by the domestic price at the partner country $w^{D\prime}$ and the barrier factor ${\mu}$ under the currency exchange factor $\nu$.
The barrier factor $\mu$ captures varius factors such as insurance, freight, miscellaneous tax, and tariff factors.
For further convenience, we may decompose ${\mu}$ into the tariff factor $1+\tau$, where $\tau$ represents the tariff rate, and other factors which we denote by $\rho$, as follows:
\begin{align}
w^P = \nu{\mu}w^{D\prime} = \nu (1+\tau)\rho w^{D\prime}
\label{thirteen}
\end{align}
As we monitor $\nu$ and ${\mu}$ for the two states, $w^P$ can be evaluated accordingly, i.e.,
\begin{align}
&w_<^P = \nu_< \cdot {\mu}_< \cdot w_<^{D\prime}
&w_>^P = \nu_> \cdot {\mu}_> \cdot w_>^{D\prime}
\label{fourteen}
\end{align}

\subsection{Multifactor CES Aggregator}
Production of industry $j$ (index omitted) is assumed to be carried out under a constant returns multifactor CES (constant elasticity of substitution) whose unit cost function can be described in the following form:
\begin{align}
w^D = t^{-1} \left( \sum_{i=0}^n \lambda_i \left( w_i^C \right)^{1-\sigma} \right)^{\frac{1}{1-\sigma}}
\label{fifteen}
\end{align}
where $\lambda_{i} \in (0,1)$ and $\sigma$ are the share parameter for the $i$th input and the multifactor CES elasticity of substitution, respectively, while
$t$ denotes the productivity level.
While $w^D$ is observable, $w_i^C$ depends on (\ref{one}) via $w^D$ and $w^F$, which are statistically observable, and the calibrated parameters $\alpha$ and $\varepsilon$.

We note below that $\sigma$ and $t$ can be estimated by regression, for each industrial sector.
The cost share of the $i$th input $s_i$ may be represented according to Shephard's lemma by differentiating (\ref{fifteen}) as follows:
\begin{align}
s_i = \frac{\partial w^D}{\partial w_i^C} \frac{w_i^C}{w^D} = \lambda_i t^{-1}\left(
\frac{w_i^C}{w^D} \right)^{1-\sigma}
\label{sixteen}
\end{align}
By taking the logarithm of both sides, we have
\begin{align}
\ln s_i = \ln \lambda_i -(1-\sigma) \ln t + (1-\sigma) \left( \ln {w_i^C} - \ln{w^D} \right)
\label{seventeen}
\end{align}
Thus, the difference in (\ref{seventeen}) between two temporally distant states, i.e., reference and current, is given by the following formula:
\begin{align}
\Delta \ln s_i = -(1-\sigma) \Delta \ln t + (1-\sigma) \left( \Delta \ln {w_i^C} - \Delta \ln {w^D} \right) 
\label{eighteen}
\end{align}
Note, if $\sigma$ and $t$ are estimated by the slope and the intercept of (\ref{eighteen}), $\lambda_i$ will be determined by (\ref{sixteen}).

\section{Measurements}
\subsection{Armington Elasticities}
A set of linked input--output tables includes sectoral transactions in both nominal and real terms.
Hence, such a set of tables provides temporally distant observations of cost shares and prices (as indexes) for all factor inputs (and outputs).
In this study, we use the 1995--2000--2005 linked input--output tables for both Japan \citep{miac} and Korea \citep{bok}, and we chose the year 2000 for reference and 2005 for current states.
In order to calibrate macro elasticity $\varepsilon_j$ on two state observations using (\ref{five}),  we standardize all prices at the \textit{current} state and evaluate the reference state prices by the current-standardized price index (the \textit{inflator}), which we denote by $q$.
Specifically, we use the following terms for calibrating the parameters:
\begin{align*}
&\left( w_<^D, w_>^D \right) =\left( q^D, 1 \right)
&\left( w_<^F, w_>^F \right) =\left( q^F, 1 \right)
\end{align*}
The parameters of the macro aggregator are thus evaluated by the following formulae, based on (\ref{five}) and (\ref{sss}):
\begin{align*}
&\alpha = s_>^D
&\varepsilon = 1+\frac{\Delta \ln s^D -\Delta \ln s^F}{\ln q^D - \ln q^F}
\end{align*}

In order to evaluate micro elasticities, we need reference and current observations of the partner country and the ROW market shares $(s_<^P, s_>^P)$ within the foreign factor inputs.
To this end, we use the 6-digit HS trade data of the UN Comtrade database \citep{comtrade}, spanning 6,376 goods, converted into the linked input--output sector classification\footnote{
Subsequent analysis will be confined to traded goods (products) while excluding services, due to data availability.
} in order to obtain the market share of the partner country with respect to that of the ROW in two periods (2000 and 2005).
Further, in order 
to calibrate the parameters of the micro aggregators, we need to specify the in-bound prices of the partner country's commodities as noted in (\ref{fourteen}).
That is, we need the inflator $q^P$, while $q^F$ is observable in the linked input--output tables.
\begin{align*}
&\left( w_<^F, w_>^F \right) =\left( q^F, 1 \right)
&\left( w_<^P, w_>^P \right) =\left( q^P, 1 \right)
\end{align*}
Therefore, we use the exchange rate that properly scales the two countries' price indexes.
Specifically, (\ref{fourteen}) must be replaced by the following identities:
\begin{align}
&q^P = \nu_< \cdot {\mu}_< \cdot q^{D\prime}
&1 = \nu_> \cdot {\mu}_> \cdot 1
\label{twenty}
\end{align}
Note that since $\nu_<{\mu}_< = (\nu_</\nu_>)({\mu}_</{\mu}_>)$, according to (\ref{twenty}), we may use current standardized index numbers for reference currency exchange factor $\nu_<$ as well as for the reference barrier factor $\mu_<$.
In this way, we evaluate the in-bound partner country's commodity inflator $q^P$ by way of an inflator of the commodity produced inside the partner country $q^{D\prime}$.
Then, according to (\ref{ten}) and (\ref{eleven}), the parameters of micro aggregator are determined by the following equations:
\begin{align*}
&\beta = s_>^P
&\eta = 1 - \frac{\Delta \ln s^P}{\ln q^F - \ln q^P}
\end{align*}

\begin{figure}[t!]
\includegraphics[width=.495\textwidth]{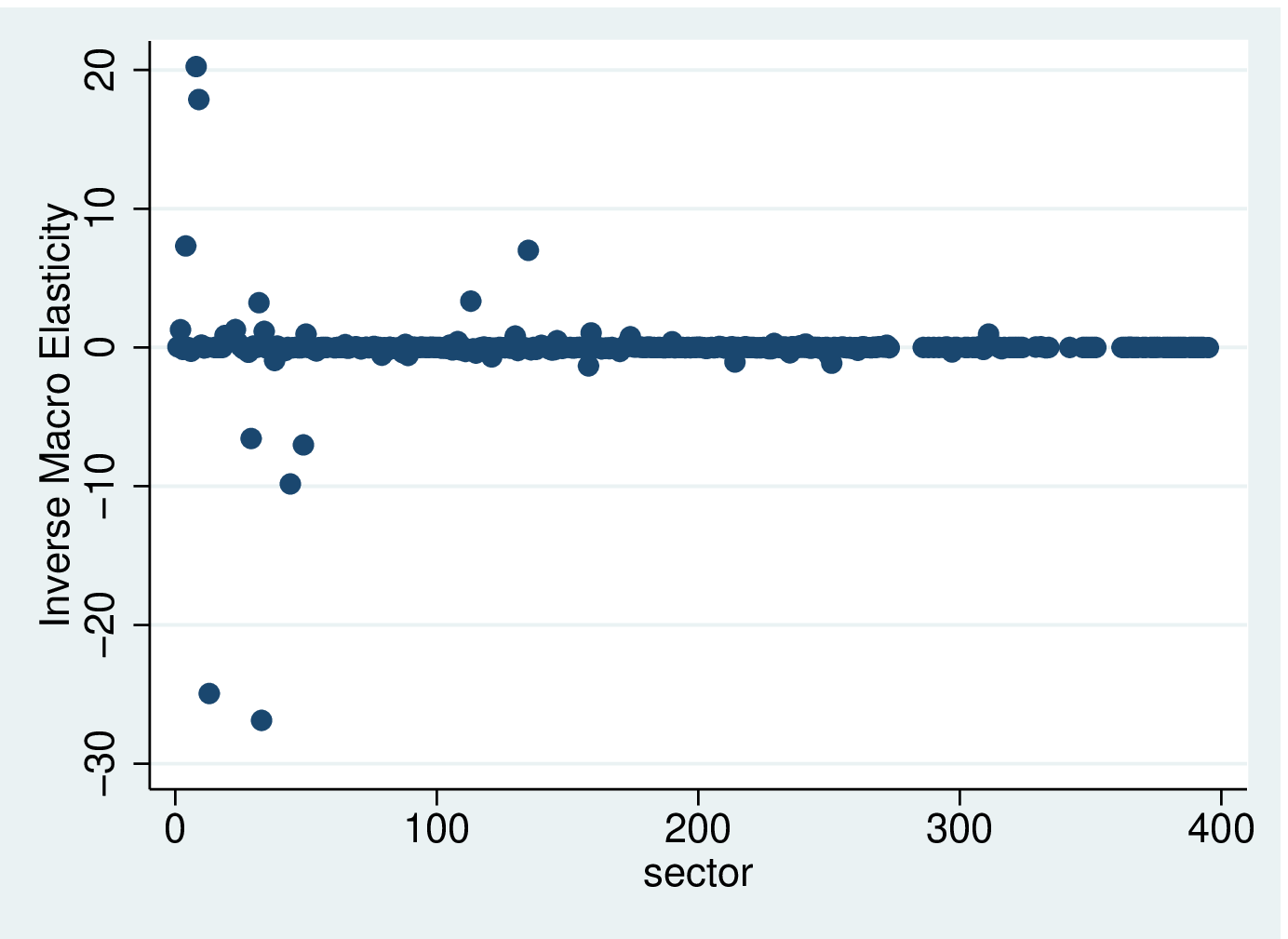}
\includegraphics[width=.495\textwidth]{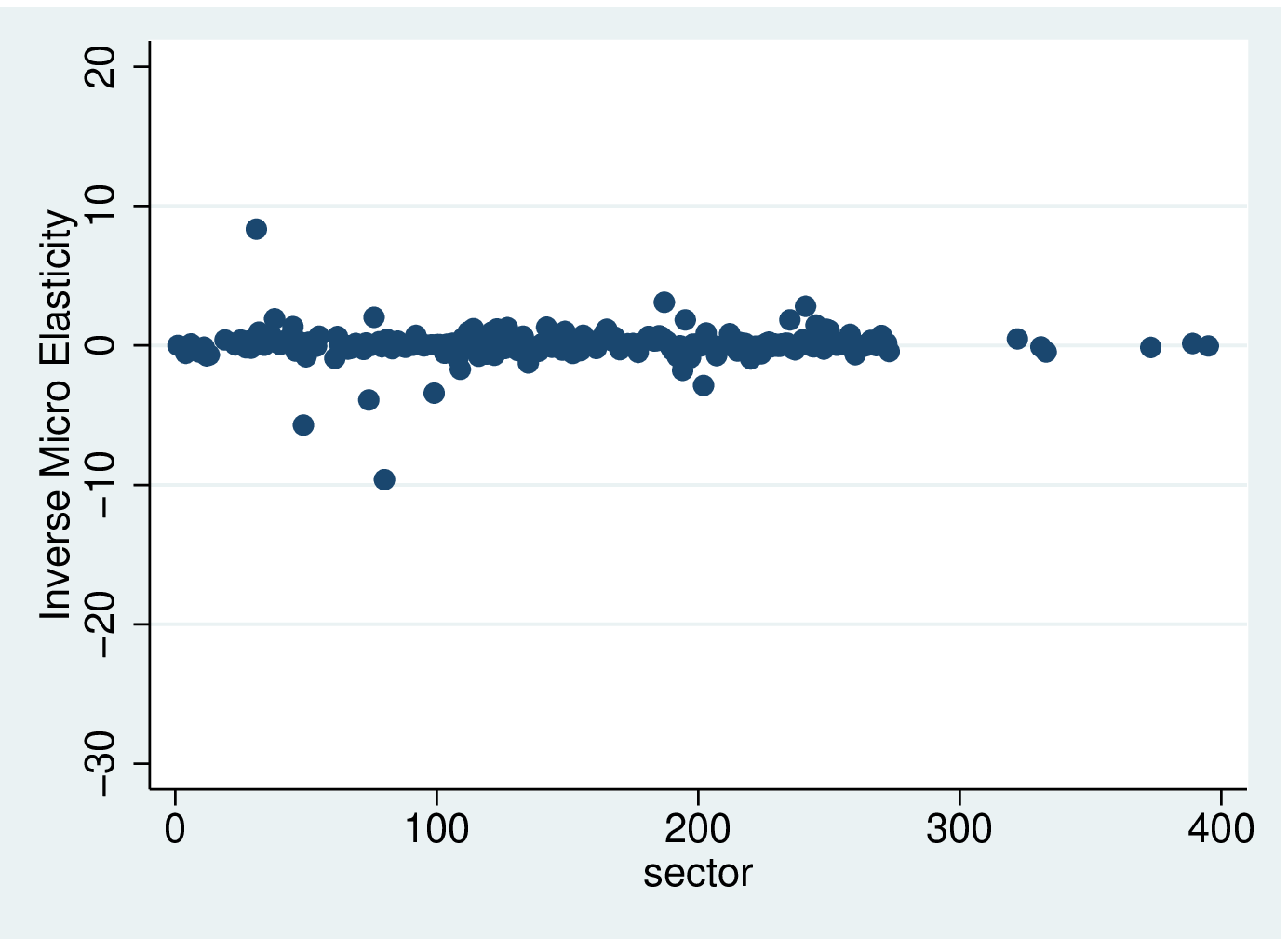}
\caption{Inverse macro Armington elasticity $\varepsilon_j^{-1}$ for Japan (left) and inverse micro Armington elasticity $\eta_j^{-1}$ for Japan against Korea (right).
}
\label{japan}      
\end{figure}
\begin{figure}[t!]
\includegraphics[width=.495\textwidth]{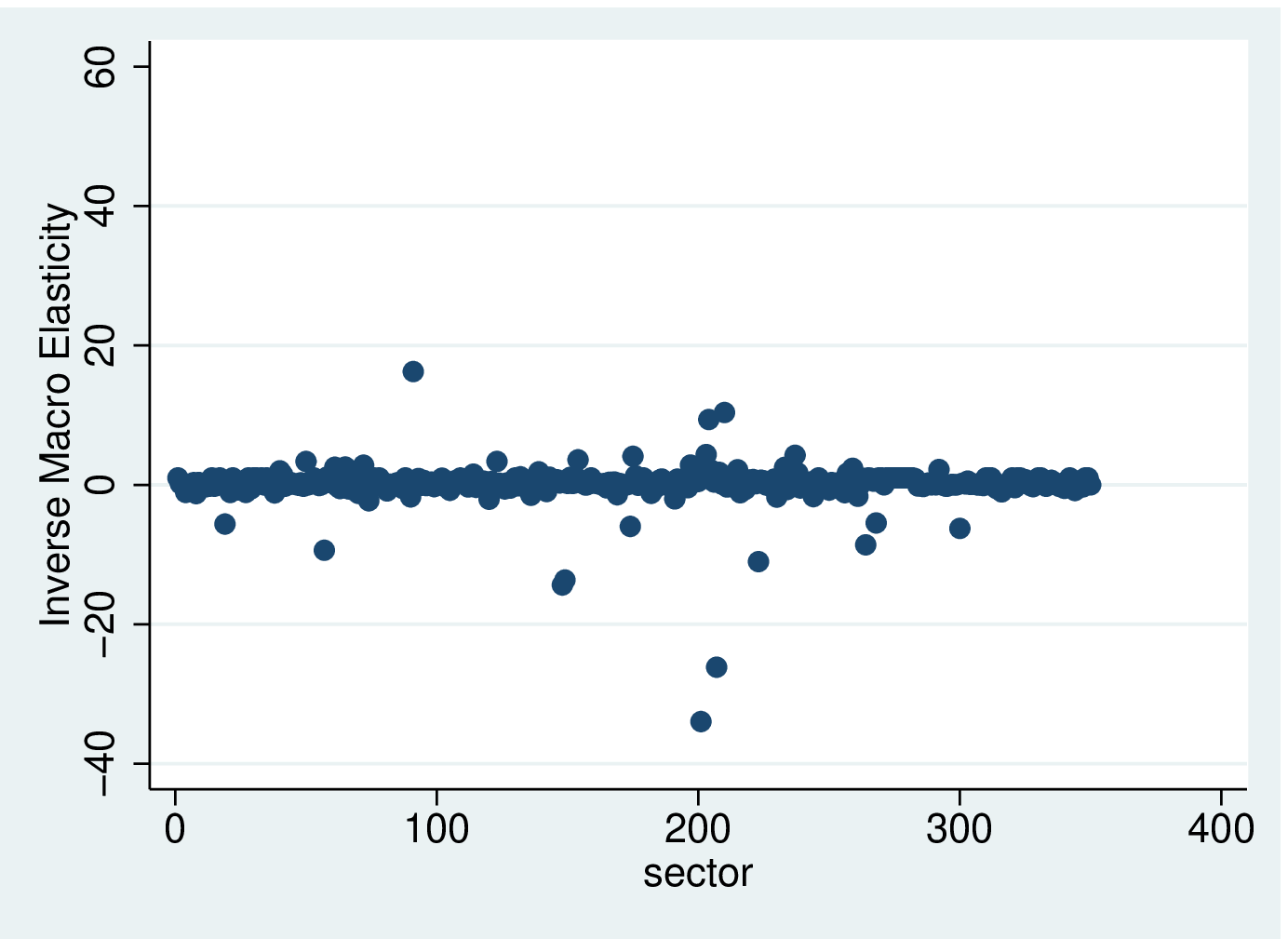}
\includegraphics[width=.495\textwidth]{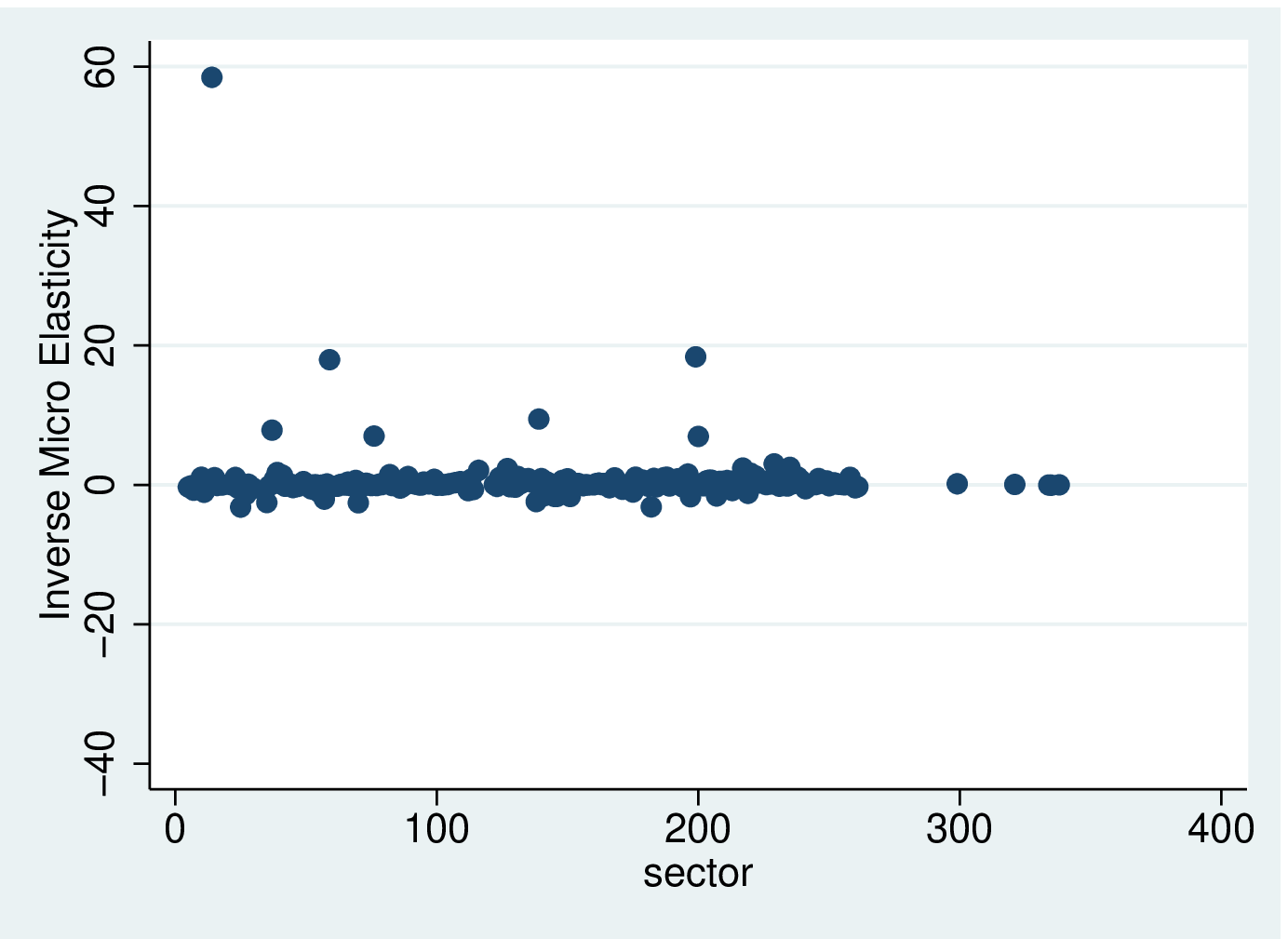}
\caption{Inverse macro Armington elasticity $\varepsilon_j^{-1}$ for Korea (left) and inverse micro Armington elasticity $\eta_j^{-1}$ for Korea against Japan (right).
}
\label{korea}       
\end{figure}

In Fig. \ref{japan}, we display the two-point calibrated macro and micro Armington elasticities of 395 sectors for Japan.
Note that the sectors are ordered according to Colin Clark's Three-sector theory, namely, 
$j=1$--27 are primary, $j=28$--294 are secondary, and $j=295$--395 are tertiary sectors.
The figure shows the reciprocals since the calibrated elasticities were very large and diverse. 
Overall, we have very large macro Armington elasticities, meaning that the domestic and imported commodities are (almost complete) substitutes, while
some of the imported commodities of the primary sectors show some extent of complementarity.
On the other hand, Japan's micro Armington elasticities relative to Korean products are relatively small, meaning that the Korean-made commodities are somehow different from those of the rest of the importing countries, for Japan.
In Fig. \ref{korea}, we display the two-point calibrated macro and micro Armington elasticities of 350 sectors for Korea.
In this case, sectors are primary for $j = 1$--28, secondary for $j = 29$--282, and tertiary for $j=283$--350.
The figure shows the reciprocals since the calibrated elasticities were very large.
Overall, Korea's macro and micro Armington elasticities are both smaller than those of Japan.
This means that Korean industries perceive foreign-made inputs to be somehow different from Korean-made inputs.
 
\subsection{Multifactor CES Elasticities}
We estimate multifactor CES elasticities for all production sectors according to the regression equation (\ref{eighteen}).
However, in this case, we must measure $\Delta \ln w_i^C$ between current and reference states in advance, using the macro aggregator (\ref{one}) whose parameters are measured via the two-point calibration method presented previously.
The reference and current compound prices evaluated with respect to the price indexes (inflators) used for domestic and foreign commodities are as follows: 
\begin{align*}
&\left( w_{<}^C, w_{>}^C \right) = \left( q^C, 1 \right)
& q^C = \left( 
\lambda \left( q^D \right)^{1-\varepsilon} 
+
(1-\lambda) \left( q^F \right)^{1-\varepsilon} 
\right)^{\frac{1}{1-\varepsilon}}
\end{align*}
Using these values, we estimate $\sigma$ via (\ref{eighteen}).
Specifically, $1-\sigma$ is estimated by the slope of the following linear regression equation:
\begin{align}
\Delta \ln s_i = - (1-\sigma) \Delta \ln t + (1-\sigma) \left( \ln q^D - \ln q_i^C \right) + u_i
\label{twentytwo}
\end{align}
where $s_i$ is the cost share of input $i$ for the concerned industrial sector whose reference and current values are both available in a set of linked input--output tables\footnote{Notice that an input--output coefficient of input $i$ for output $j$ represents the cost share of factor $i$ for industry $j$.} and $u_i$ is the disturbance term.
Further, note that growth of productivity, i.e., $\Delta \ln t$, is estimable from the intercept of the regression line, although that analysis is beyond the purpose of this study.

\begin{figure}[t!]
\includegraphics[width=.495\textwidth]{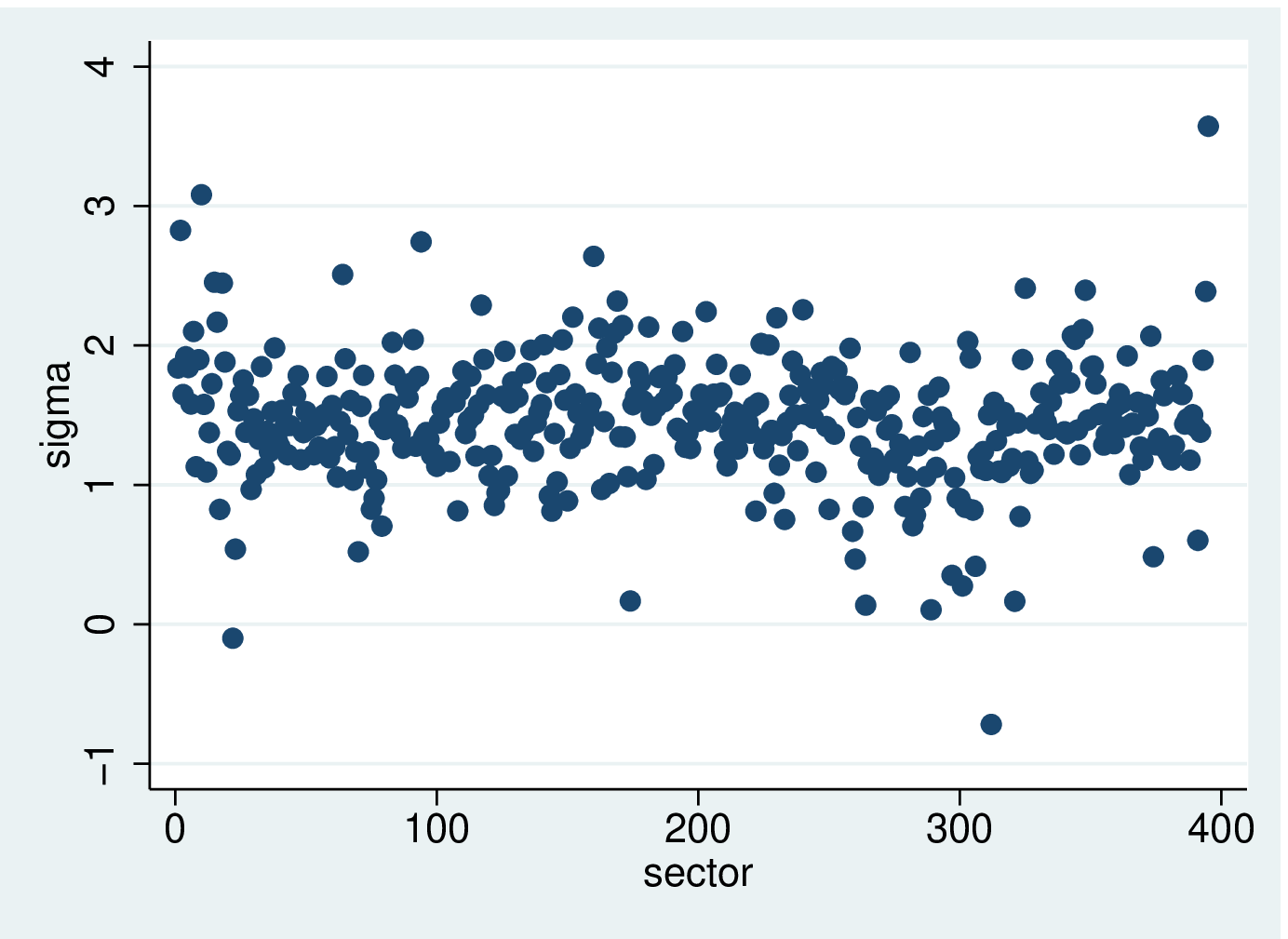}
\includegraphics[width=.495\textwidth]{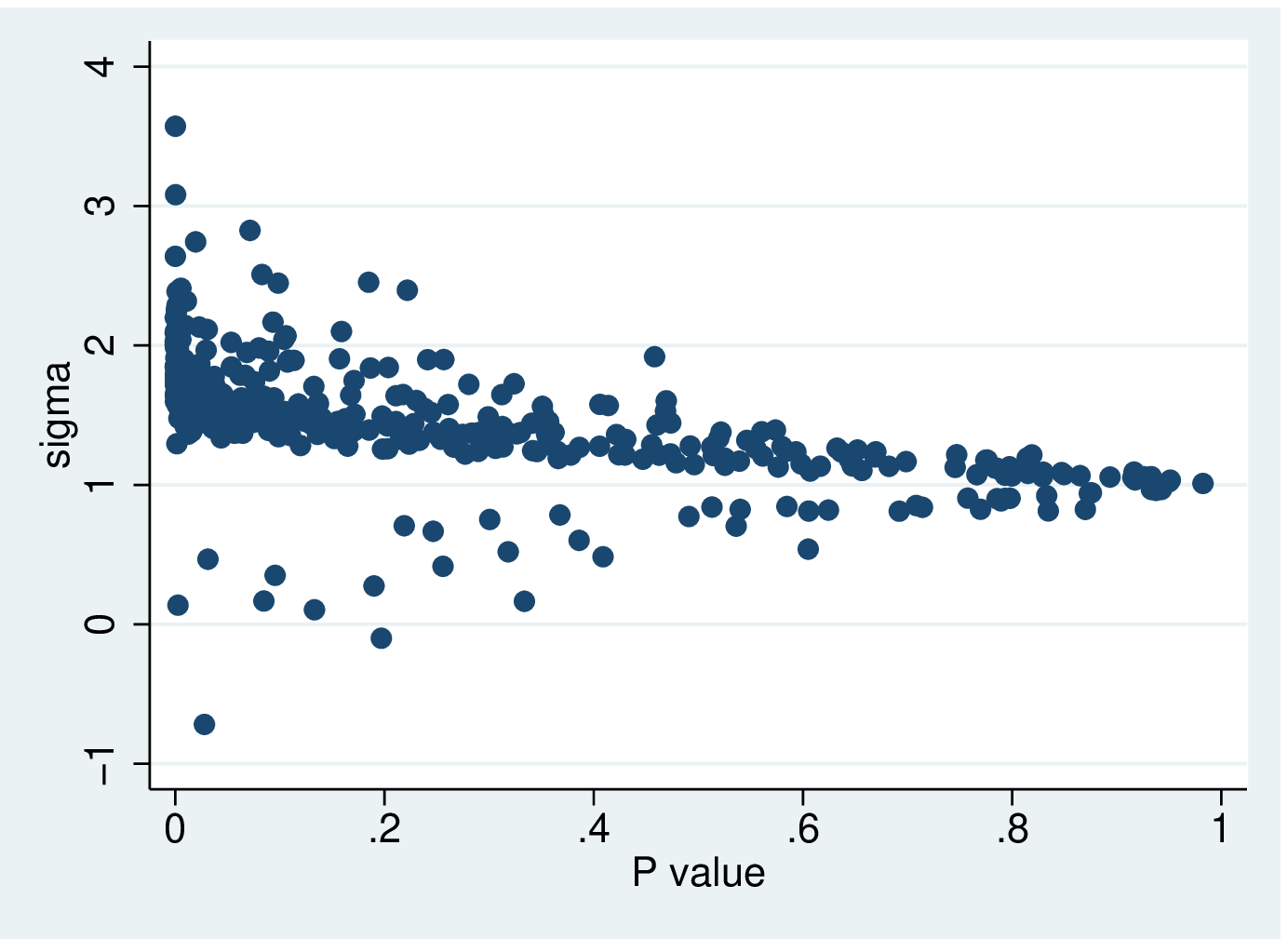}
\caption{Multifactor CES elasticities $\sigma_j$ estimated for Japan.
}
\label{japanCES}       
\end{figure}

\begin{figure}[t!]
\includegraphics[width=.495\textwidth]{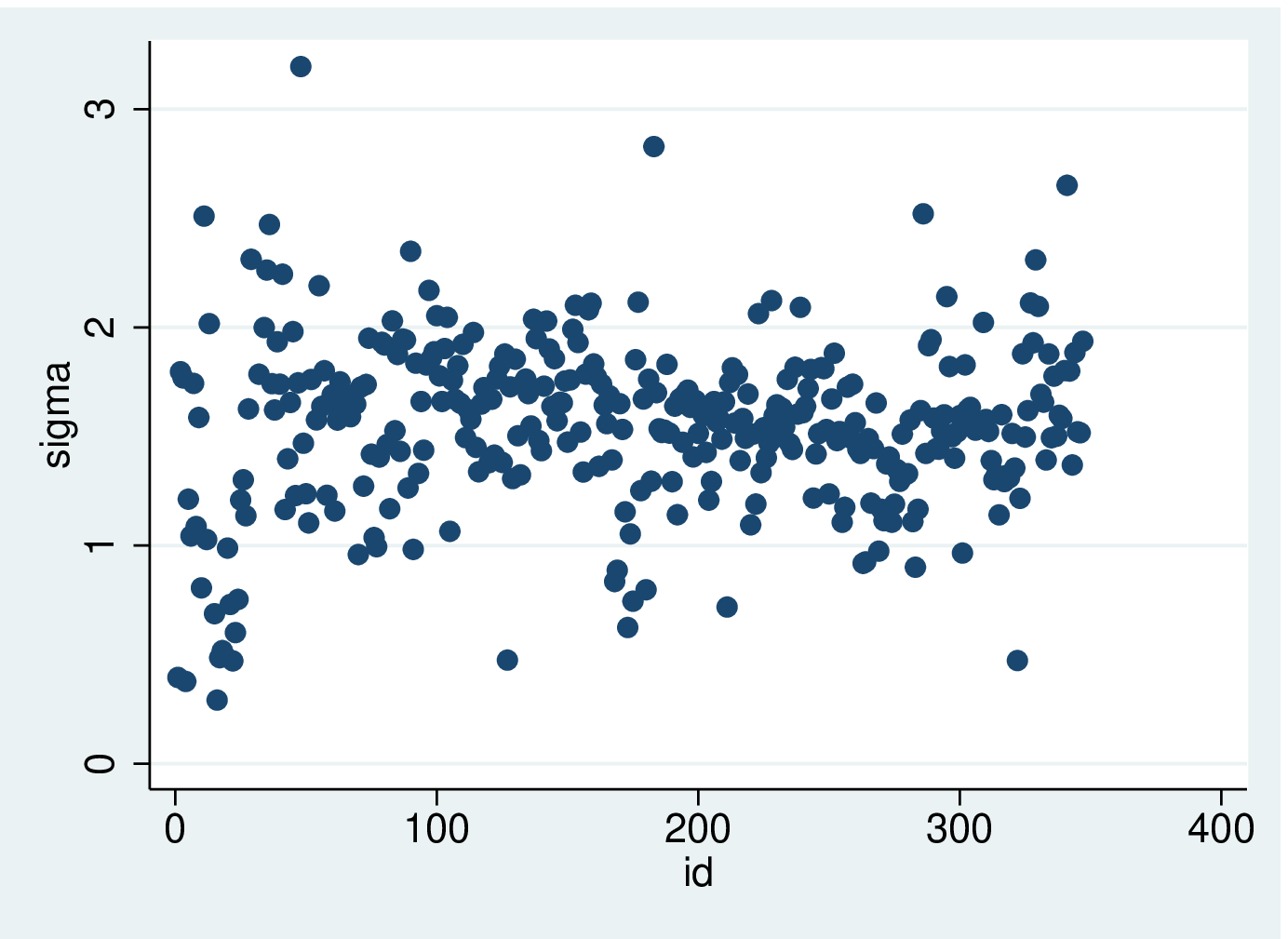}
\includegraphics[width=.495\textwidth]{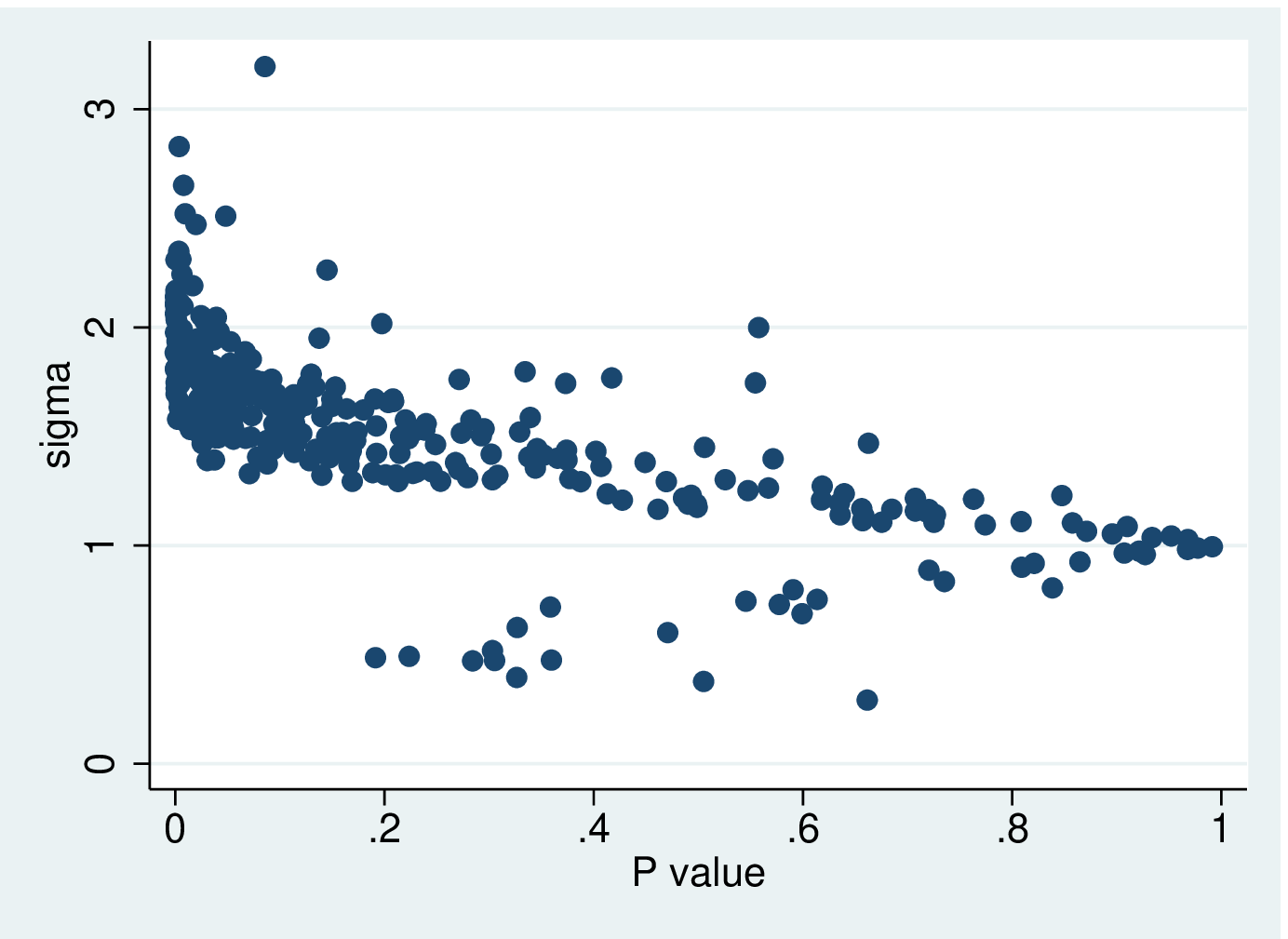}
\caption{Multifactor CES elasticities $\sigma_j$ estimated for Korea.
}
\label{koreaCES}       
\end{figure}

We must note that linked input--output tables do not provide price indexes for the primary input (comprising labor and capital), which we aggregate as a single input in this study.
To address this, we use the quality-adjusted price indexes of labor and capital compiled by \citet{jip} for Japan and by \citet{kip} for Korea for the corresponding periods in order to inflate the value added observed in nominal values.
In Fig. \ref{japanCES},  we report the estimated multifactor CES elasticities $\sigma$ for all sectors (left) with the corresponding statistical significances (right) for Japan. 
Fig. \ref{koreaCES} is the equivalent figure for Korea.
Further, we shall note that the average of the estimated elasticities (ignoring statistical significance) is 1.46 for Japan and 1.53 for Korea, and these values are almost identical to those estimated by using $q_i^D$ instead of $q_i^C$ in regression equation (\ref{twentytwo}) as reported in \citet{knn}.

\section{Bilateral Equilibrium}
\subsection{Model Integration}
In this section, we construct a bilateral multisectoral general equilibrium model that reflects all measured elasticities for the two countries. 
Let us first focus on one country's general equilibrium state of multisectoral production.
We shall calibrate the share parameters at the current state in order to examine various policy shifts (such as tariff elimination) on the basis of the current state.
As we have previously arranged that all current prices be the basis of price standardization, we may calibrate the share parameters $\lambda_i$ at the current state where the productivity is standardized at unity $t=1$, according to (\ref{sixteen}):
\begin{align}
&\lambda_i = a_i
\label{twentythree}
\end{align}
Here, $a_i$ is the current state input--output coefficient (i.e., cost share) of input $i$ for the industry (output) concerned, and thus, $\sum_{i=0}^n a_i =1$.
We may express the system of unit cost functions (\ref{fifteen}) as
\begin{align*}
w_1^D &= \left( 
a_{01} ( w_0^C )^{1-\sigma_1}
+a_{11} ( w_1^C )^{1-\sigma_1}
+ \cdots
+a_{n1} ( w_n^C )^{1-\sigma_1}
\right)^{\frac{1}{1-\sigma_1}}
\\
w_2^D &= \left( 
a_{02} ( w_0^C )^{1-\sigma_2}
+a_{12} ( w_1^C )^{1-\sigma_2}
+ \cdots
+a_{n2} ( w_n^C )^{1-\sigma_2}
\right)^{\frac{1}{1-\sigma_2}}
\\
&~~\vdots \\
w_n^D &= \left( 
a_{0n} ( w_0^C )^{1-\sigma_n}
+a_{1n} ( w_1^C )^{1-\sigma_n}
+ \cdots
+a_{nn} ( w_n^C )^{1-\sigma_n}
\right)^{\frac{1}{1-\sigma_n}}
\end{align*}
or more concisely as 
\begin{align}
\mathbf{w}^D = {H} \left( \mathbf{w}^C, w_0^C 
\right)
\label{twentyfour}
\end{align}

The model for both countries according to the multifactor CES aggregator (\ref{fifteen}), the macro aggregator (\ref{one}), and the micro aggregator (\ref{six}), can be expressed as follows, where $J$ and $K$ indicate Japan and Korea, respectively:
\begin{align}
\mathbf{w}_J^D &= {H}_J \left( \mathbf{w}_J^C \right)
&\mathbf{w}_K^D &= {H}_K \left( \mathbf{w}_K^C \right) 
\label{twofive}
\\
\mathbf{w}_J^C &= {U}_J \left( \mathbf{w}_J^D, \mathbf{w}_J^F \right)
&\mathbf{w}_K^C &= {U}_K \left( \mathbf{w}_K^D, \mathbf{w}_K^F \right)
\\
\mathbf{w}_J^F &= {V}_J \left( \mathbf{w}_J^P \right)
&\mathbf{w}_K^F &= {V}_K \left( \mathbf{w}_K^P \right) \label{twoseven}
\end{align}
Note that we eliminate $w_0^C$ from the multifactor CES aggregator since it is fixed as constant and 
$\mathbf{w}^R$ from the micro aggregators as we assume that ROW import prices are invariable (under the small-country assumption).

In order to close (integrate) the model, we must introduce a weighted converter that connects the foreign sector with the domestic sector classifications in terms of 6-digit HS transactions.
Specifically, a sector-HS converter $z_{jk}$ that assigns a sectoral commodity $j$ to an HS item $k$ has the following form:
\begin{align*}
z_{jk} = \frac{x_{jk}}{\sum_{k \in j} x_{jk}}
\end{align*}
where $x_{jk}$ represents the amount of import of HS item $k$ that belongs to sector $j$.
As we represent Japan's sector-HS converter by matrix $\mathbf{z}_J$ and Korea's sector-HS converter by $\mathbf{z}_K$, Korea's 350 sectors can be converted into Japan's 395 sectors by $\mathbf{z}_K\mathbf{z}_J^{\intercal}$, and likewise Japan's sectors can be converted into Korea's by $\mathbf{z}_J\mathbf{z}_K^{\intercal}$, where $\intercal$ indicates transposition.
Thereupon, we introduce the following identities, according to (\ref{thirteen}):
\begin{align}
&\mathbf{w}_J^P = \mathbf{w}_K^{D}\mathbf{z}_K\mathbf{z}_J^{\intercal}\left< \boldsymbol{\nu}_J\right>\left<\boldsymbol{\mu}_J \right> 
&\mathbf{w}_K^P = \mathbf{w}_J^{D}\mathbf{z}_J\mathbf{z}_K^{\intercal}\left< \boldsymbol{\nu}_K\right>\left<\boldsymbol{\mu}_K \right> 
\label{twoeight}
\end{align}
where angle brackets indicate diagonalization.
Additionally, we know that $\nu \mu=1$ from (\ref{twenty}) at the current state.
Hence, we know that the equilibrium solution to the bilateral integrated price system (\ref{twofive})--(\ref{twoeight}) at the current state is unity for all, i.e., $\mathbf{w}_J^D = \mathbf{w}_J^C= \mathbf{w}_J^F= \mathbf{w}_J^P=\mathbf{1}$ in terms of Japan's sectors and $\mathbf{w}_K^D = \mathbf{w}_K^C= \mathbf{w}_K^F= \mathbf{w}_K^P=\mathbf{1}$  in terms of Korea's sectors.

\subsection{Tariff Elimination}
\begin{figure}[t!]
\includegraphics[width=.495\textwidth]{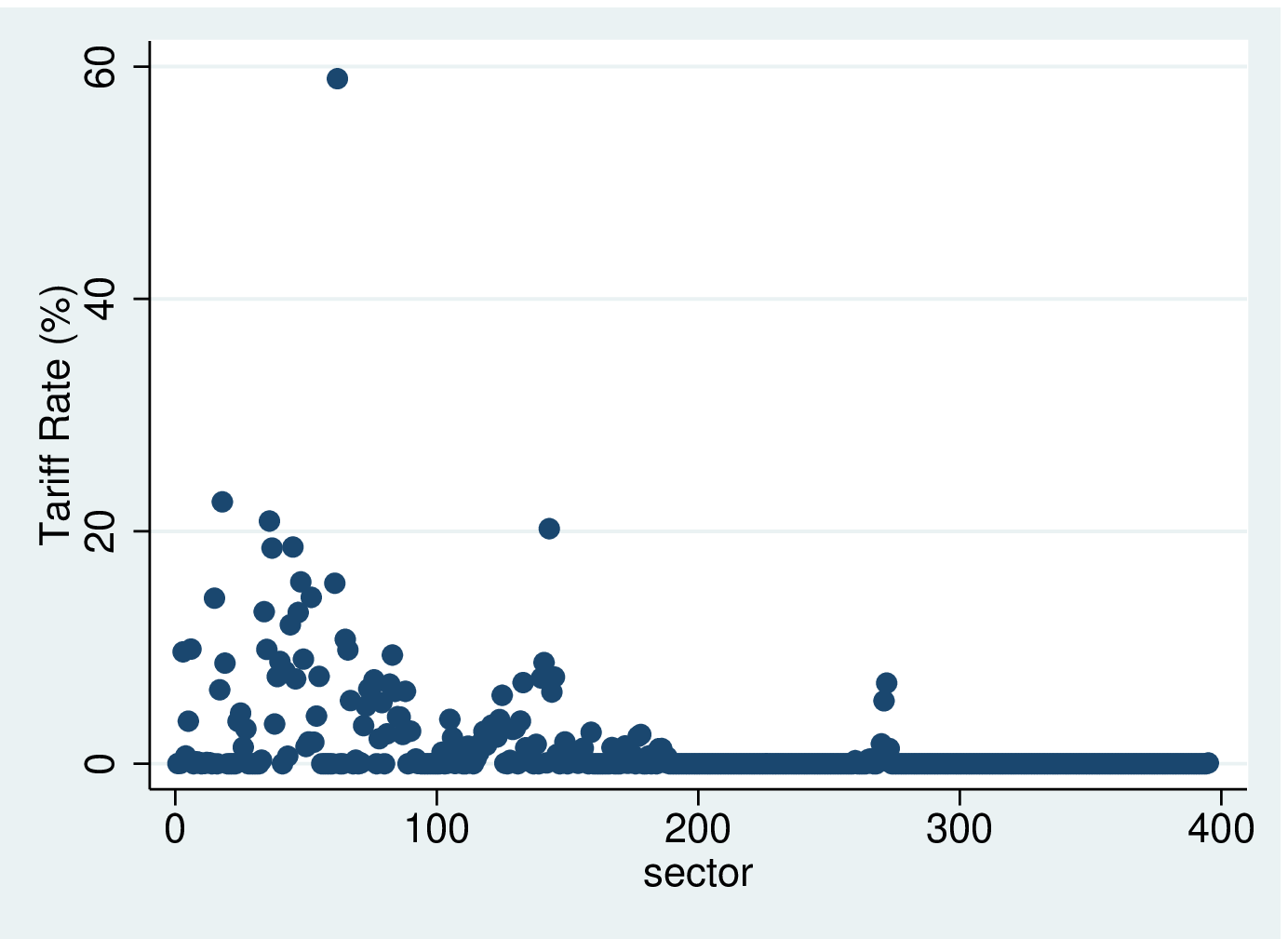}
\includegraphics[width=.495\textwidth]{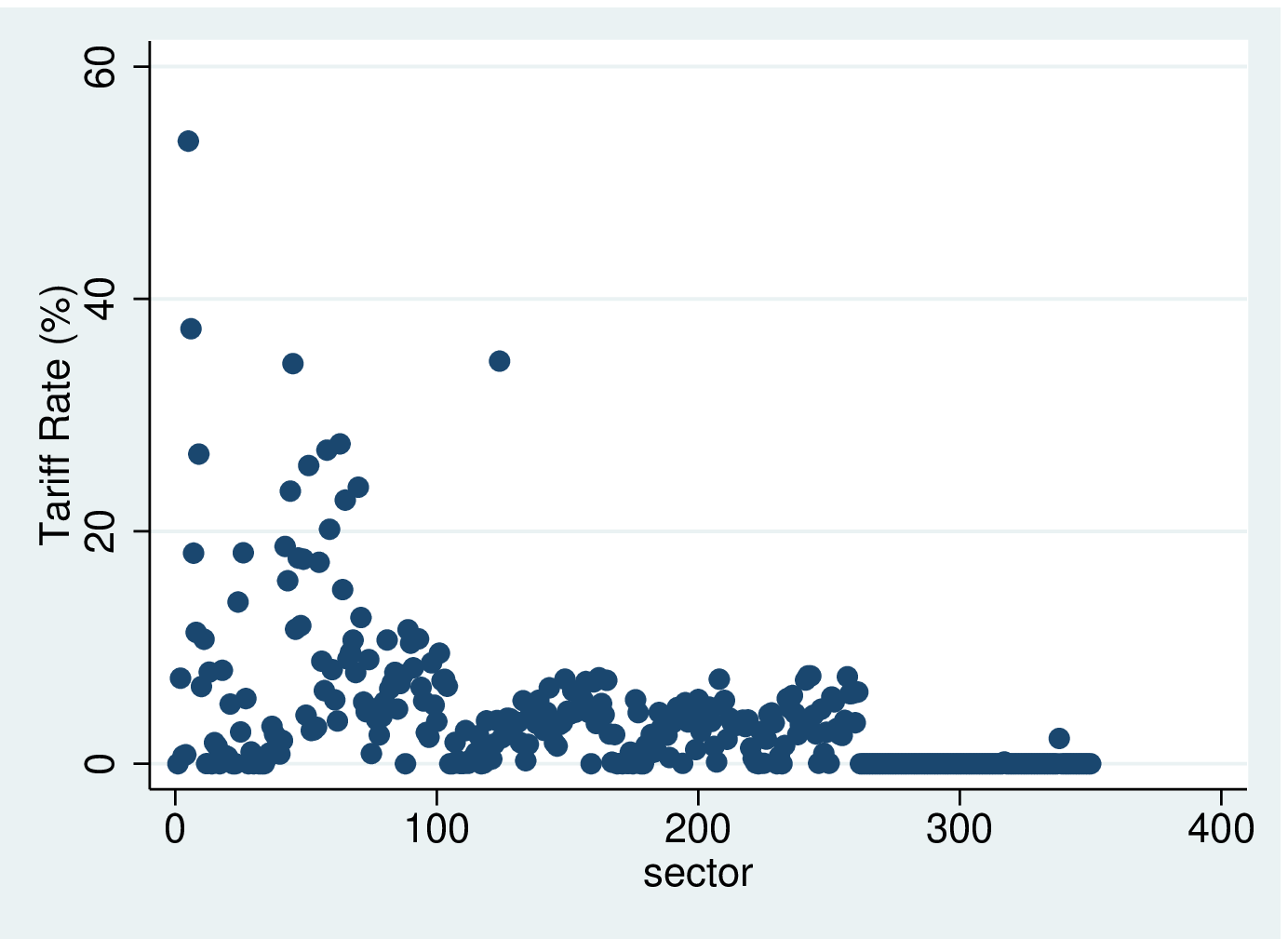}
\caption{Tariff rates $\tau$ for Japan against Korea (left) and for Korea against Japan (right).
}
\label{tariff}       
\end{figure}

We first calculate the equilibrium price when all tariffs that levied against the partner country's commodities in both countries were eliminated.
For the purpose we specify the tariff rates levied at the current state and thus we used the UNCTAD Trade Analysis Information System \citep{trains} database.
Specifically, we used the tariff rates evaluated by way of customs duties-imported values that were converted into ratios and distributed over the linked input--output product classifications.
In Fig. \ref{tariff} we display the estimated tariff rates levied against the partner country's commodities for all sectoral commodities, for both countries.
Note that ``Refined sake'' (59.0\%) and ``Beef cattle'' (22.5\%) were among the higher tariff rate commodities in Japan against Korea, whereas ``Vegetables'' (53.6\%) and ``Fruits'' (37.4\%) were among the higher tariff rate commodities in Korea against Japan.

Let us now consider what happens if the tariff between the two countires were entirely eliminated over the current state.
In that event the ex ante barrier factor $\mu^*$ will equal $\rho$ instead of $\rho (1+\tau)$, in regard to (\ref{thirteen}).\footnote{Here, we are assuming the exchange factor $\nu$ to be constant.}
Thus, because $\nu{\mu}=1$ at the current state according to (\ref{twenty}), ${\mu}^*$ must be evaluated as follows:
\begin{align*}
\nu {\mu}^* = \nu\rho 
=\frac{\nu{\mu}}{1+\tau}
=\frac{1}{1+\tau}
\equiv \theta
\end{align*}
and hence, we must modify (\ref{twoeight}) when evaluating tariff-eliminated bilateral general equilibrium prices, as follows:
\begin{align}
&\mathbf{w}_J^P = \mathbf{w}_K^{D}\mathbf{z}_K\mathbf{z}_J^{\intercal}\left< \boldsymbol{\theta}_J \right> 
&\mathbf{w}_K^P = \mathbf{w}_J^{D}\mathbf{z}_J\mathbf{z}_K^{\intercal}\left< \boldsymbol{\theta}_K \right> 
\label{thirty}
\end{align}

Hereafter let us denote by $\boldsymbol{\pi}$ the tariff-eliminated bilateral general equilibrium prices.
That is, 
\begin{align*}
\boldsymbol{\pi} = \left( \boldsymbol{\pi}_J^D, \boldsymbol{\pi}_J^C, \boldsymbol{\pi}_J^F, \boldsymbol{\pi}_J^P, \boldsymbol{\pi}_K^D, \boldsymbol{\pi}_K^C, \boldsymbol{\pi}_K^F, \boldsymbol{\pi}_K^P \right)
\end{align*}
More specifically $\boldsymbol{\pi}$ is the fix point of the mapping $G: \mathbb{R}^{4(n_J+n_K)} \to \mathbb{R}^{4(n_J+n_K)}$ which comprises of  the functions (\ref{twofive}--\ref{twoseven}) and (\ref{thirty}) i.e.,\footnote{The dimension of the sectors are $n_J=395$ for Japan and $n_K=350$ for Korea.} 
\begin{align}
\boldsymbol{\pi} = G\left( \boldsymbol{\pi} \right)
\label{threetwo}
\end{align}
Note that $G$ is a concave and monotone increasing mapping because CES aggregators $H$, ${U}$ and ${V}$ are all concave functions, and linear functions (\ref{thirty}) are also concave (although not strictly concave).
Thus, $G$ becomes a contraction mapping and we may solve (\ref{threetwo}) for the fixed point by recursive means \citep[see e.g.,][]{kennan, kras} from arbitrary initial guess such as $\mathbf{1}$ i.e.,
\begin{align*}
\boldsymbol{\pi} = \lim_{k \to \infty} G^k \left( \mathbf{1} \right) 
= 
G\left(\cdots 
G\left(
G\left(G\left( \mathbf{1} \right) \right)
\right)
\cdots \right) 
\end{align*}

\subsection{Prospective Analysis}

Since we know by the Shephard's lemma that the factor input can be obtained by differentiating the unit cost function, inputs in physical units per physical unit output for all sectors, or the physical input--output coefficient matrix, can be obtained as the gradient of (\ref{twentyfour}) i.e., 
\begin{align*}
\nabla \mathbf{w}^D 
=
\begin{bmatrix}
\frac{\partial {H}_1\left( \mathbf{w}^C, w_0^C\right)}{\partial w_0^C} & \frac{\partial {H}_2\left( \mathbf{w}^C, w_0^C\right)}{\partial w_0^C} & \cdots & \frac{\partial {H}_n\left( \mathbf{w}^C, w_0^C\right)}{\partial w_0^C}  \\
\frac{\partial {H}_1\left( \mathbf{w}^C, w_0^C\right)}{\partial w_1^C} & \frac{\partial {H}_2\left( \mathbf{w}^C, w_0^C\right)}{\partial w_1^C} & \cdots & \frac{\partial {H}_n\left( \mathbf{w}^C, w_0^C\right)}{\partial w_1^C}  \\
\vdots & \vdots & \ddots & \vdots  \\
\frac{\partial {H}_1\left( \mathbf{w}^C, w_0^C\right)}{\partial w_n^C} & \frac{\partial {H}_2\left( \mathbf{w}^C, w_0^C\right)}{\partial w_n^C} & \cdots & \frac{\partial {H}_n\left( \mathbf{w}^C, w_0^C\right)}{\partial w_n^C}  
\end{bmatrix}
=
\begin{bmatrix}
\nabla_0 H
\left( \mathbf{w}^C, w_0^C\right)  \\
\nabla H
\left( \mathbf{w}^C, w_0^C\right)
\end{bmatrix}
\end{align*}
where $\nabla_0{H}$ is an $n$ row vector, while $\nabla{H}$ is an $n\times n$ matrix. 
For later convenience, let us use the following terms to indicate monetary input--output coefficient matrices for current and posterior (with tariff elimination) states.
\begin{align*}
\textstyle
1 \nabla_0 H \left( \mathbf{1}, 1 \right) \left< \mathbf{1} \right>^{-1}
&\equiv \mathbf{{a}}_0
& \textstyle
\pi_0^C
\nabla_0 H \left( \boldsymbol{\pi}^C, \pi_0^C \right) \textstyle \left< \boldsymbol{\pi}^C \right>^{-1} 
&\equiv \mathbf{\tilde{a}}_0 
\\
\left< \mathbf{1} \right> \nabla H \left( \mathbf{1}, 1 \right)  \left< \mathbf{1} \right>^{-1}
&\equiv {\mathbf{{A}}}
&\textstyle
\left< 
\boldsymbol{\pi}^C 
\right> 
\nabla H \left( \boldsymbol{\pi}^C, \pi_0^C \right) \left< \boldsymbol{\pi}^C \right>^{-1} 
&\equiv \mathbf{\tilde{A}}
\end{align*}
Note that we set $\pi_0^C =1$ as we take the primary input $i=0$, which is not produced industrially, as the num{\'e}raire good.
Also, $\mathbf{a}_0$ and $\mathbf{A}$ are the current state (observed) value-added and input--output coefficients, respectively.

Below is the commodity balance in monetary terms:
\begin{align}
\mathbf{x} = \mathbf{A} \mathbf{x} + \mathbf{y} + \mathbf{e} - \mathbf{m}
\label{threefour}
\end{align}
where, $\mathbf{x}$ denotes domestic output, $\mathbf{y}$ denotes domestic final demand, $\mathbf{e}$ denotes export, $\mathbf{m}$ denotes import, all in column vectors of monetary terms, while $\mathbf{A}\mathbf{x}$ represents the intermediate demand.
Here, we may recall that we have obtained the current state foreign share of a commodity $s_>^F=1-\alpha$, by the amount of import $m$ (i.e., an element of $\mathbf{m}$ whose index is omitted) 
and the domestic total demand, i.e.,
\begin{align*}
{m} = 
(1-\alpha)
\left( y + \sum_{j=1}^n a_j x_j \right)
\end{align*}
For further conveniece let us define $s = s_>^F=1-\alpha$ and endogenize import with respect to the domestic total demand as follows:
\begin{align*}
\mathbf{m} = \left<\mathbf{s}\right> [\mathbf{A}\mathbf{x} + \mathbf{y}]
\end{align*}
Further, we may recall that the import from the partner country $\mathbf{m}^P$ can be replicated by the current share of the partner country's commodity $s_>^P$, which we hereafter denote $s^P$ for convenience, as follows: 
\begin{align*}
\textstyle
\mathbf{m}^P = \left< \mathbf{s}^P \right> \mathbf{m}=\left< \mathbf{s}^P \right>\left< \mathbf{s} \right> \left[ \mathbf{Ax} + \mathbf{y} \right]
\end{align*}

Displayed below is the commodity balance of the posterior state:
\begin{align}
\tilde{\mathbf{x}}=
\tilde{\mathbf{A}} \tilde{\mathbf{x}} + \tilde{\mathbf{y}} + \left( \mathbf{e}^W + \tilde{\mathbf{e}}^P \right) 
- \left( \tilde{\mathbf{m}}^W + \tilde{\mathbf{m}}^P \right)
\label{epcomb}
\end{align}
The posterior state values are distinguished by tildes.   
We assume that imports and exports are subject to change due to tariff elimination, except for the exports to the ROW.
Notice that imports from the partner and the ROW are assumed to be proportional to the total domestic demands in the following manner:
\begin{align}
&\textstyle \tilde{\mathbf{m}}^P = \left< \tilde{\mathbf{s}}^P \right> \left< \tilde{\mathbf{s}} \right> \left[ \tilde{\mathbf{A}} \tilde{\mathbf{x}} + \tilde{\mathbf{y}} \right]=\tilde{\mathbf{e}}^{P\prime}
&\textstyle \tilde{\mathbf{m}}^W = \left[ \mathbf{I} - \left< {\mathbf{s}}^P \right> \right] \left< {\mathbf{s}} \right> \left[ \tilde{\mathbf{A}} \tilde{\mathbf{x}} + \tilde{\mathbf{y}} \right]
\label{mm}
\end{align}
As indicated above, the export against the partner country is determined by the import from the partner's partner country.\footnote{Here, a prime is used to indicate the partner country's export against its partner country.}
The import coefficients are determined by (\ref{two}) and (\ref{nine}) as follows:
\begin{align*}
&\tilde{s}=(1-\alpha) \left( \frac{\pi^F}{\pi^C} \right)^{1-\varepsilon}
&\tilde{s}^P=\beta \left( \frac{\pi^P}{\pi^F} \right)^{1-\eta}
\end{align*}
The posterior value added (external inputs) total can be evaluated by the import endogenized model in regard to the posterior commodity balance equation (\ref{epcomb}):\footnote{This model is otherwise called the Chenery--Moses type or the competitive import model.}
\begin{align}
\tilde{\mathbf{a}}_0
\tilde{\mathbf{x}}
=
\tilde{\mathbf{a}}_0
\left[ \mathbf{I}-\left[ \mathbf{I} - \left< {\mathbf{s}}^* \right> \right] \tilde{\mathbf{A}} \right]^{-1} \left[ \left[ \mathbf{I} - \left< {\mathbf{s}}^* \right> \right] \tilde{\mathbf{y}} + \mathbf{e}^W
+\tilde{\mathbf{e}}^P
 \right]
 \label{endo}
\end{align}
where, the import coefficient $\mathbf{s}^*$ is specified as follows, according to (\ref{mm}):
\begin{align*}
\textstyle
\left< {\mathbf{s}}^* \right>
=\left< \tilde{\mathbf{s}}^P \right> \left< \tilde{\mathbf{s}} \right>
+
\left[ \mathbf{I} - \left< {\mathbf{s}}^P \right> \right] \left< {\mathbf{s}} \right>
\end{align*}

We assume an economy to maximize its final demand given the external inputs total, and to this end the compensation of increased exports against the partner country can be spent for whatever commodity demanded.
We incorporate such external inputs into the domestic production in such a way that the external inputs (value added) total is fortified.\footnote{It may be more natural to incorporate export compensation into imports; but this option was not adopted on the ground that imports are endoginized with respect to domestic final demand alone as specified in (\ref{mm}).} 
In particular, we are to find a scalar $\delta$ of the following problem that maximizes the total ex ante value of the current-proportioned final demand i.e., $\tilde{\mathbf{y}}=\left< \boldsymbol{\pi}^{D}\right> \mathbf{y} \delta$, given the ex ante 
total value added (\ref{endo}), which is limited to the sum of the locally existing primary factor $\ell$ $(= \mathbf{a}_0 \mathbf{x})$, and the compensation of exports against the partner country, i.e.,
\begin{align}
\textstyle
\max_{\delta} \, \mathbf{1} \tilde{\mathbf{y}}=\mathbf{1} \left< \boldsymbol{\pi}^D \right>  \mathbf{y} \delta \text{~~~s.t.~~~~~} \tilde{\mathbf{a}}_0 
\tilde{\mathbf{x}}
\leq \ell + \mathbf{1} \left( \tilde{\mathbf{e}}^P -\mathbf{e}^P \right)
\label{max}
\end{align}
Note that the solution of (\ref{max}) determines the posterior total domestic demands and thus the imports from the partner country which, in turn, determines the compesation of exports against the partner's partner country via (\ref{mm}) that must enter into the constraint of the parter country's problem.
In other words, (\ref{max}) must be solved recurrsively for both countries under the condition given by the partner country.
\begin{figure}[tp!]
\includegraphics[width=0.495\textwidth]{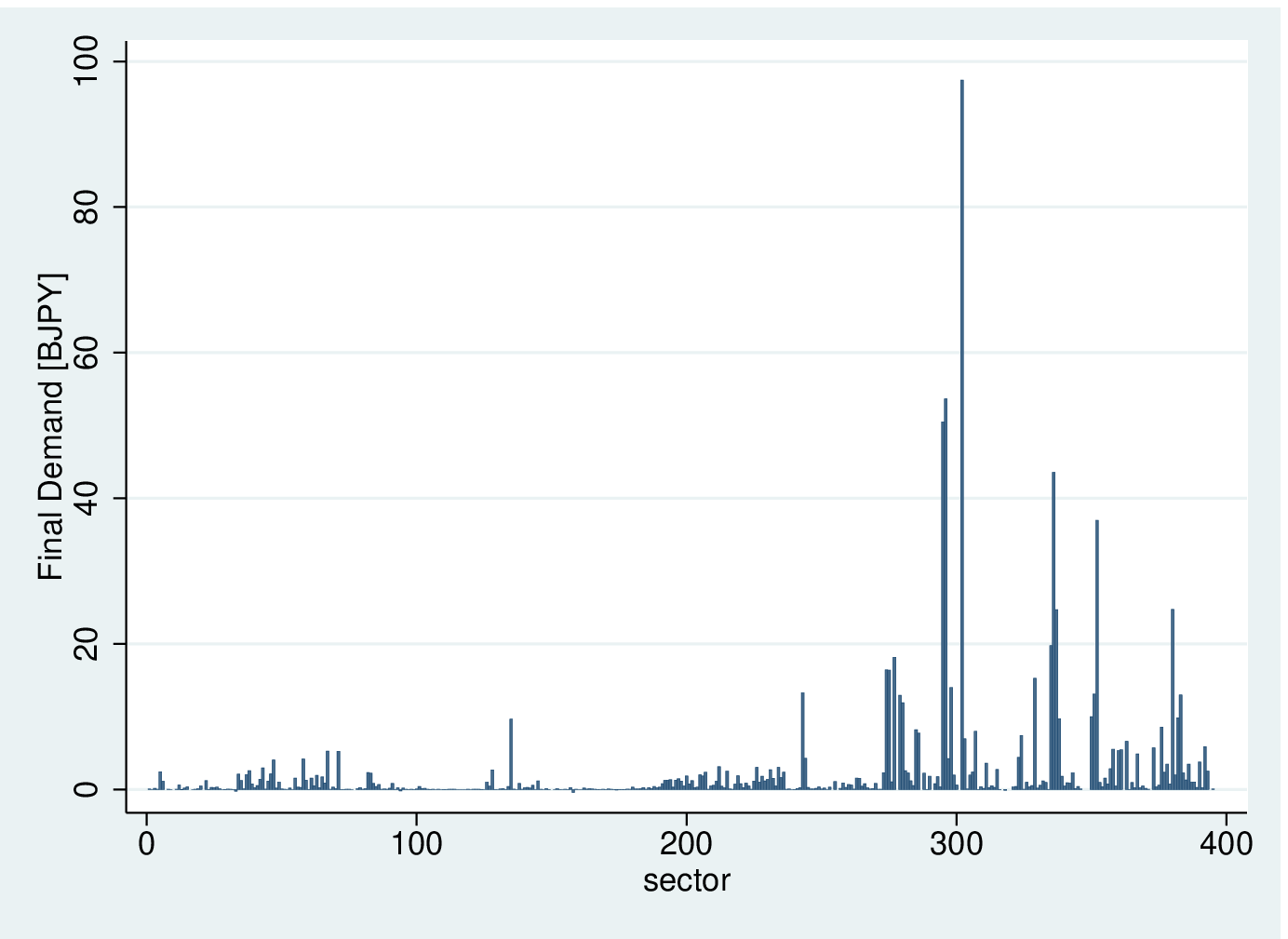}
\includegraphics[width=0.495\textwidth]{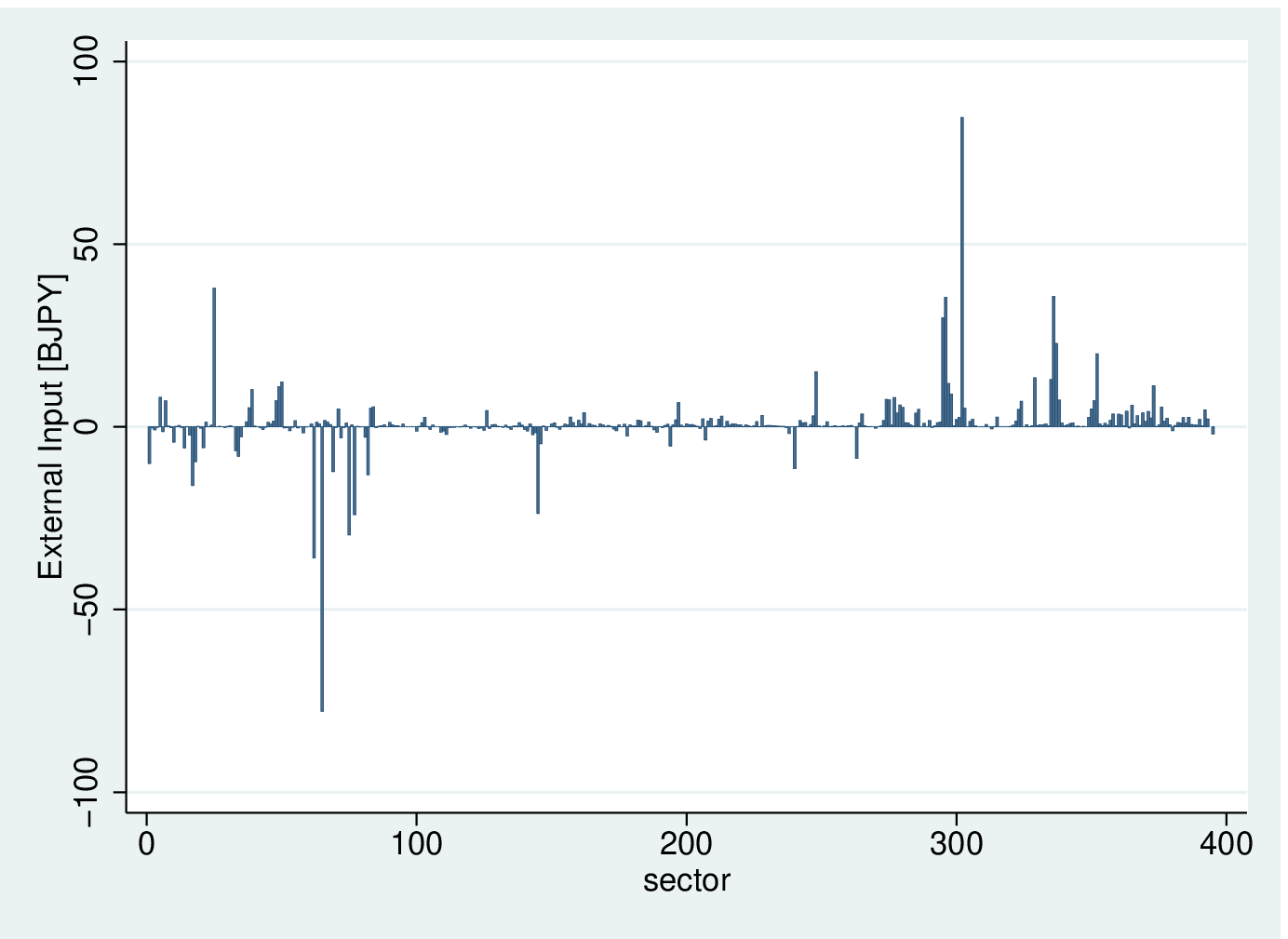}
\caption{Maximized increment of current-proportioned final demand $\tilde{\mathbf{y}} -\mathbf{y}$ (left) 
and the corresponding redistribution of the external inputs $\tilde{\mathbf{a}}_0 \tilde{\mathbf{x}} -{\mathbf{a}}_0 \mathbf{x}$ (right) for Japan.}
\label{jp}       
\end{figure}

\begin{figure}[t!]
\includegraphics[width=0.495\textwidth]{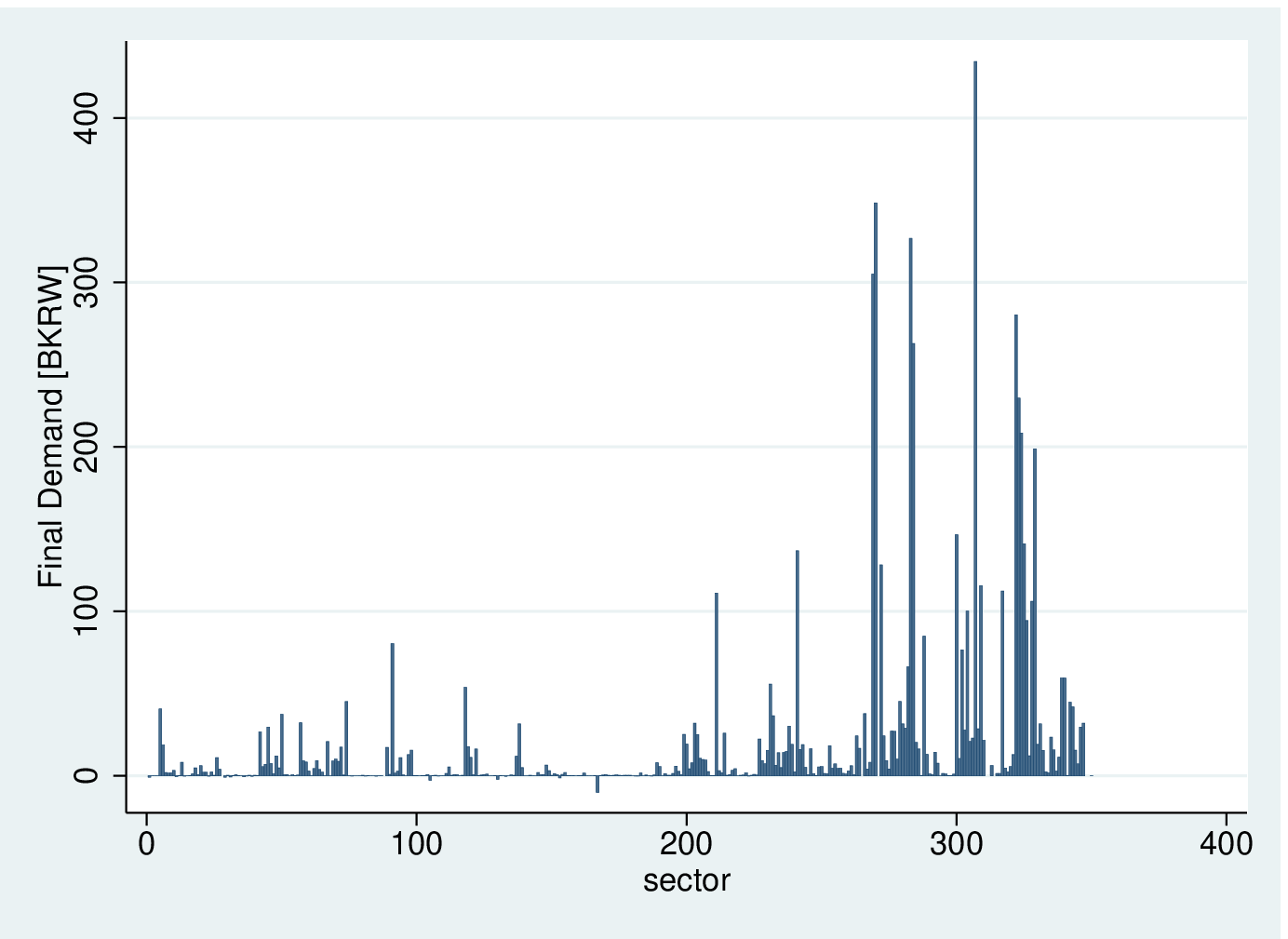}
\includegraphics[width=0.495\textwidth]{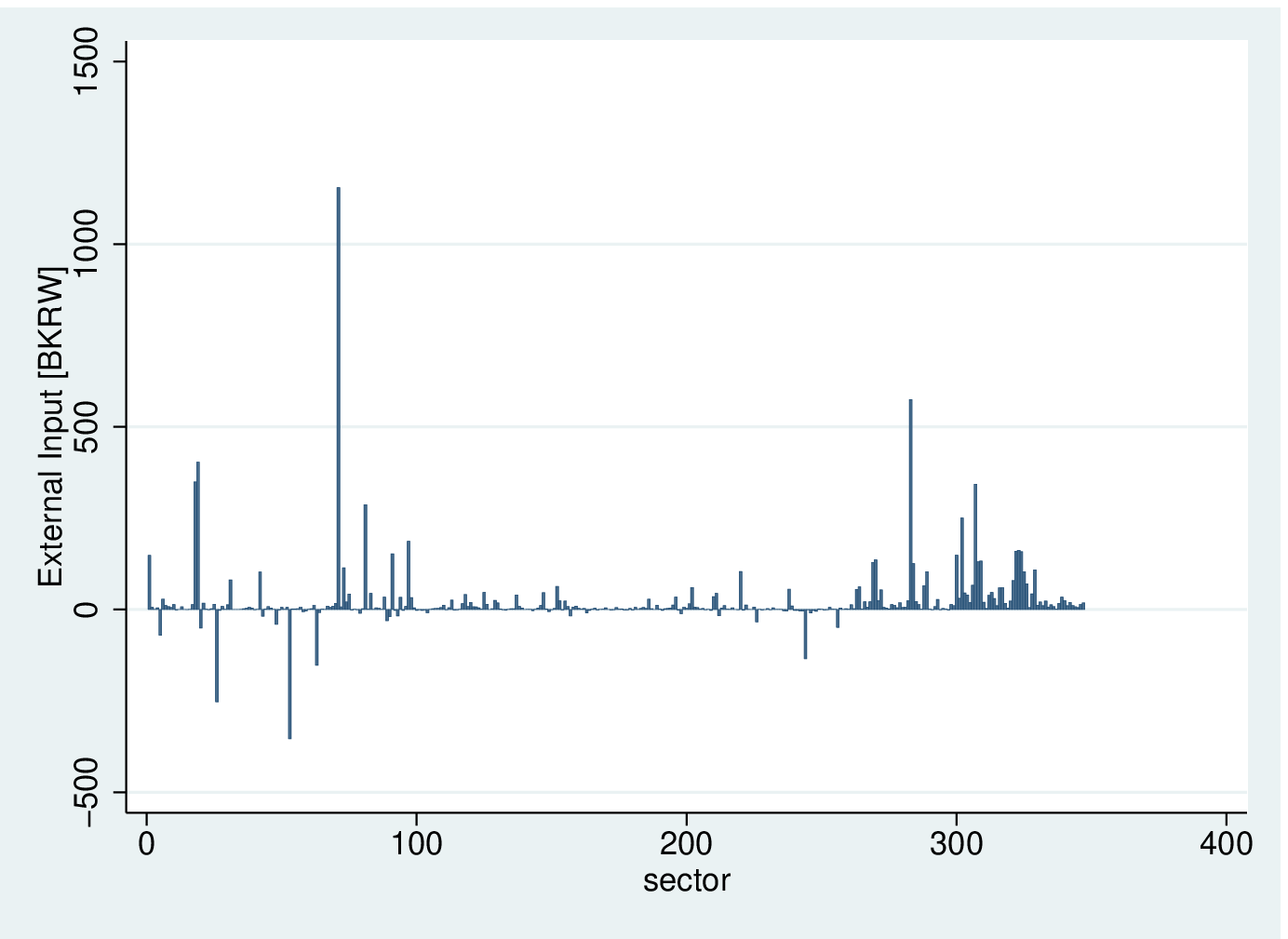}
\caption{Maximized increment of current-proportioned final demand $\tilde{\mathbf{y}} -\mathbf{y}$ (left) 
and the corresponding redistribution of the external inputs $\tilde{\mathbf{a}}_0 \tilde{\mathbf{x}} -{\mathbf{a}}_0 \mathbf{x}$ (right) for Korea.}
\label{kr}       
\end{figure}
\begin{table}[tp!]
\caption{Prospective analysis of tariff elimination between Japan and Korea.} \label{aaa}
\begin{tabular}{rrrrr}
\toprule
 & \multicolumn{2}{c}{Japan} & \multicolumn{2}{c}{Korea} \\
\cmidrule(l{.75em}r{.75em}){2-3}\cmidrule(l{.75em}r{.75em}){4-5}
 & BJPY & (BKRW) & BKRW & (BJPY) \\
\midrule
Gross Domestic Product (GDP) & 505,269 & 
 &  851,982 & 
 \\
$\Delta$ Final Demand $\Delta y$ & 853
 & 7,924 & 6,309 & 679 \\
$\Delta$ Import from Partner $\Delta m^P$ & 803 & 7,459 & 3,338 & 359
\\
$\Delta$ Export to Partner $\Delta e^{P}$ & 359 & 3,338 & 7,459 & 803 \\
\bottomrule
\end{tabular}
\end{table}

Figs. \ref{jp} and \ref{kr} illustrate the increments of maximized current-proportioned final demand i.e., $\tilde{\mathbf{y}} -\mathbf{y}$ and the corresponding redistribution of the external inputs i.e., $\tilde{\mathbf{a}}_0 \tilde{\mathbf{x}} -{\mathbf{a}}_0 \mathbf{x}$
for Japan and Korea, respectively, under the tariff elimination between the two countries.\footnote{As regards (\ref{max}), the increment total of external inputs must be equal to the increment total of export compensation from the partner country i.e., $\mathbf{1} (\tilde{\mathbf{a}}_0 \tilde{\mathbf{x}} -{\mathbf{a}}_0 \mathbf{x}) = \mathbf{1} (\tilde{\mathbf{e}}^P - \mathbf{e}^P )$.}
Notice that BJPY stands for billion Japanese yens and BKRW for billion Korean wons.
The total effects are summarized in Table \ref{aaa}.
The net benefit (in terms of gained final demand $\Delta y$) of tariff elimination is about 0.17\% of the current GDP (853 BJPY) for Japan, whereas it is about 0.74\% of the current GDP (6,309 BKRW) for Korea.
As regards the redistribution of the external inputs, 
current-proportioned final demand maximization suggests that sectors such as
$j=302$ (House rent),
$j=25$ (Fisheries),
$j=352$ (Medical service (medical corporations, etc.)),
and 
$j=329$ (Information services)
must be reinforced, and curtailed in sectors such as
$j=65$ (Other liquors),
$j=75$ (Woolen fabrics, hemp fabrics and other fabrics),
$j=145$ (Miscellaneous leather products),
and
$j=17$ (Hogs), for Japan.
On the other hand, preferable policy for Korea is to reinforce in sectors such as
$j=71$ (Other liquors),
$j=283$ (Wholesale and Retail trade),
$j=19$ (Pigs),
and 
$j=18$ (Beef cattle), and to curtail in sectors such as
$j=53$ (Raw sugar),
$j=26$ (Fishing),
$j=63$ (Canned or cured fruits and vegetables),
and
$j=17$ (Motor vehicle engines, chassis, bodies and parts), for Korea.

Table \ref{aaa} also displays the import change from the partner country $\Delta m^P$ defined as below, which by definition equals the export change of the partner country against its partner country $\Delta e^{P\prime}$, or more specifically,
\begin{align*}
\Delta m^P_J 
&= \mathbf{1} \left( \tilde{\mathbf{m}}^P_J - {\mathbf{m}}^P_J \right)
= \mathbf{1} \Delta{\mathbf{m}}^{P}_J 
=\Delta e^{P}_K 
\\
\Delta m^{P}_K
&= \mathbf{1} \left( \tilde{\mathbf{m}}^{P}_K - {\mathbf{m}}^{P}_K \right)
= \mathbf{1} \Delta{\mathbf{m}}^{P}_K 
=\Delta e^{P}_J 
\end{align*}
Thus, we may look into the \textit{net export} $\Delta f$ against the partner country as follows:
\begin{align*}
&\Delta e_J^P - \Delta m_J^P = \Delta m_K^P - \Delta e_K^P = \Delta e_J^P - \Delta e_K^P= \Delta m_K^P - \Delta m_J^P
\equiv
\Delta f_{JK}
\\
&\Delta e_K^P - \Delta m_K^P = \Delta m_J^P - \Delta e_M^P= \Delta e_K^P - \Delta e_J^P= \Delta m_J^P - \Delta m_K^P
\equiv
\Delta f_{KJ}
\end{align*}
where naturally $\Delta f_{JK}+\Delta f_{KJ}=0$.
\begin{table}[tp]
\caption{Notable net export (top 20) from Japan to Korea $\Delta f_{JK}$.} \label{jptab}
\footnotesize
\begin{tabular}{rrr}
\toprule
sector/commodity & BJPY  & BKRW \\
\midrule
Fisheries	&	48	&	446	\\
Sugar	&	38	&	350	\\
Motor vehicle parts and accessories	&	34	&	321	\\
Salted, dried or smoked seafood	&	32	&	298	\\
Bottled or canned vegetables and fruits	&	29	&	272	\\
Preserved agricultural foodstuffs (other than bottled or canned)	&	28	&	258	\\
Other wearing apparel and clothing accessories	&	18	&	164	\\
Machinery and equipment for construction and mining	&	16	&	148	\\
Knitted apparel	&	15	&	136	\\
Toys and games	&	12	&	110	\\
Frozen fish and shellfish	&	11	&	103	\\
Electric audio equipment	&	11	&	99	\\
Activities not elsewhere classified	&	10	&	96	\\
Rotating electrical equipment	&	8	&	76	\\
Hot rolled steel	&	8	&	76	\\
Synthetic fibers	&	7	&	65	\\
Internal combustion engines for motor vehicles and parts	&	6	&	60	\\
Machinery for service industry	&	6	&	54	\\
Miscellaneous ceramic, stone and clay products	&	5	&	48	\\
Trucks, buses and other cars	&	5	&	44	\\
\bottomrule
\end{tabular}
\end{table}
\begin{table}[tp!]
\caption{Notable net export (top 20) from Korea to Japan $\Delta f_{KJ}$.} \label{krtab}
\footnotesize
\begin{tabular}{rrr}
\toprule
sector/commodity & BKRW &BJPY  \\
\midrule
Slaughtering and meat processing	&	1,966	&	212	\\
Woolen fabrics, hemp fabrics and other fabrics	&	1,360	&	146	\\
Other liquors	&	1,350	&	145	\\
Refined sake	&	647	&	70	\\
Miscellaneous leather products	&	567	&	61	\\
Liquid crystal element	&	434	&	47	\\
Woven fabric apparel	&	322	&	35	\\
Analytical instruments, testing machine, measuring instruments	&	206	&	22	\\
Pumps and compressors	&	136	&	15	\\
Rolled and drawn aluminum	&	134	&	14	\\
Petrochemical aromatic products (except synthetic resin)	&	131	&	14	\\
Coal mining , crude petroleum and natural gas	&	109	&	12	\\
Petrochemical basic products	&	106	&	11	\\
Sheet glass and safety glass	&	103	&	11	\\
Metal molds	&	84	&	9	\\
Leather footwear	&	54	&	6	\\
Integrated circuits	&	53	&	6	\\
Other non-ferrous metals	&	47	&	5	\\
Fruits	&	46	&	5	\\
Petroleum refinery products (incl. greases)	&	44	&	5	\\
\bottomrule
\end{tabular}
\end{table}
In Table \ref{jptab} we display the positive entries of the net export from Japan to Korea $\Delta f_{JK}$.\footnote{The negative entries are the net import from Japan to Korea, or, the net export from Korea to Japan.}
Likewise, Table \ref{krtab} is the positive entries of the net export from Korea to Japan.

We may notice from these tables that, a lot of meat (i.e., Slaughtering and meat processing) will be exported from Korea to Japan, whereas Japan will export fish (i.e., Fisheries, Frozen fish and shellfish) to Korea, under tariff elimination.
Other notable features are that Korea will net-export petrochemical products (e.g., Petrochemical aromatic productss (except synthetic resin), Coal mining, crude petroleum and natrual gas, Petrochemical basic products, Petroleum refinery products (incl. greases), etc.) to Japan, whereas Japan will net-export mechanical and assembling products (e.g., Motor vehicle parts and accessories, Machinery and equipment for construction and mining, Electric audio equipment, Rotating electrical equipment, etc.) to Korea.

\section{Concluding Remarks}
The highlight of this study may perhaps be the discovery of a way to calibrate the parameters of a two-input CES aggregator in order that the aggregator completely replicates the observed two temporally distant shares of inputs in both monetary and physical terms.
The elasticity parameters i.e., the Armington elasticities that we obtained by way of this approach (i.e., two-point calibration), were found to be much larger than those observed in the previous studies based upon time series, implying almost complete substitutability between foreign and domestic commodities, which should not be too surprising.
We then used the Armington aggregator functions to uncover the composite price index for each commodity, which is key for modeling production activities comprising many factor inputs, including imported commodities, for each industrial sector. 

As we are concerned with multisectoral production functions of multiple (more than two) factor inputs, we estimated multifactor CES production elasticities by linearly regressing the growth of commodity-wise cost shares against the relative growths of factor prices.   
We used published statistics, namely, linked input--output tables and the UN Comtrade database, to measure all the concerned elasticities (i.e., multifactor CES production, and micro and macro Armington elasticities) for both Japan and Korea. 
The two multisectoral general equilibrium models for Japan and Korea were integrated by the bilateral trading models which reflect the trade barriers between the two countries.

Since the models presume constant returns in all activities and thus interact entirely in terms of unit costs and prices, we were able to simulate the bilateral general equilibrium consequences of eliminating tariffs between the two countries, without (physically) quantitative consideration.
The consequential social benefits and costs of tariff elimination were estimated by the amount of linear final demand that can potentially be enhanced under the projected structure for a given total primary input.
The result implies positive effects (in terms of total net benefit) for both countries, while 
considerable structural change is expected to be inevitable.

\begin{flushleft}
{\bf Conflict of Interest:} The authors declare that they have no conflict of interest.
\end{flushleft}
\bibliographystyle{spbasic_x}      
\raggedright
\bibliography{biblio}   

%
%

\end{document}